\documentclass[usenatbib]{mn2e}
\usepackage{graphicx}
\usepackage{color}

\title[Binary interactions on the calibrations of SFR]
{Binary interactions on the calibrations of star formation rate}

\author[F. Zhang, L. Li, Y. Zhang, X. Kang, Z. Han]
{Fenghui~Zhang\thanks{E-mail: gssephd@public.km.yn.cn; zhang\_fh@hotmail.com}$^{1,2}$, Lifang~Li$^{1,2}$, Yu~Zhang$^{1,2,3}$, Xiaoyu~Kang$^{1,2,3}$ and Zhanwen~Han$^{1,2}$\\
$^1$National Astronomical Observatories/Yunnan Observatory, Chinese Academy of Sciences, Kunming, 650011, China \\
$^2$Key Laboratory for the Structure and Evolution of Celestial Objects, Chinese Academy of Sciences, Kunming, 650011, China \\
$^3$Graduate University of the Chinese Academy of Science, Beijing 100049, China}

\font\hf = cmsl7 scaled \magstep 0

\begin{document}
\date{\today}
\pagerange{\pageref{firstpage}--\pageref{lastpage}}
\pubyear{2011}
\maketitle
\label{firstpage}

\begin{abstract}
Using the {\hf Yunnan} evolutionary population synthesis (EPS) models with and without binary interactions, we present the luminosity of $\rm H\alpha$ recombination line ($L_{\rm H\alpha}$), the luminosity of [OII]$\lambda$3727 forbidden-line doublet ($L_{\rm [OII]}$), the ultraviolet (UV) fluxes at 1500 and 2800\,$\rm \AA$ ($L_{i, {\rm UV}}$) and far-infrared flux ($L_{\rm FIR}$) for Burst, S0, Sa-Sd and Irr galaxies, and present the calibrations of star formation rate (SFR) in terms of these diagnostics.

By comparison, we find that binary interactions lower the SFR.vs.$L_{\rm H\alpha}$ and SFR.vs.$L_{\rm [OII]}$ conversion factors by $\sim$0.2\,dex. The main reason is that binary interactions raise the UV flux (shortward of the Lyman limit) of the stellar population (SP) in the age range 6.7$<$ log$t$/yr $<$8.4 and thus more ionizing photons are present in the nebula.
Moreover, binary interactions do not significantly vary the calibrations of SFR in terms of $L_{i, {\rm UV}}$. This is because binary interactions raise the flux at 1500\,$\rm \AA$ of the SP in the range 8.75 $<$ log$t$/yr $<$ 9.2 and the maximal difference is about 1\,dex. In addition, binary interactions have little effect on the flux at 2800\,$\rm \AA$.
At last, the calibration of SFR from $L_{\rm FIR}$ is almost unaffected by binary interactions. This is caused by the fact that binary interactions almost do not affect the bolometric magnitudes of SPs.

We also discuss the effects of initial mass function (IMF), gas-recycle assumption and EPS models (including {\hf GISSEL98, BC03, STARBURST99, PopSTAR and P\'{E}GASE} models) on these SFR calibrations.
Comparing the results by using Salpeter (S55) IMF with those by using Miller \& Scalo (MS79) IMF, we find that the SFR.vs.$L_{\rm H\alpha}$ and SFR.vs.$L_{\rm [OII]}$ conversion factors by using S55 IMF are greater by 0.4 and 0.2\,dex than those by using MS79 IMF for the {\hf Yunnan} models with and without binary interactions, respectively.
The SFR.vs.$L_{i,{\rm UV}}$ and SFR.vs.$L_{\rm FIR}$ conversion factors by using S55 IMF are larger by an amount of 0.2\,dex than the corresponding ones by using MS79 IMF.
The inclusion of gas-recycle assumption only lowers these SFR calibrations at faint SFR.
Moreover, comparing the results when using different EPS models, we find that the differences in the SFR.vs.$L_{\rm H\alpha}$ and SFR.vs.$L_{\rm [OII]}$ conversion factors reach $\sim$ 0.7 and 0.9\,dex, the difference in the SFR.vs.$L_{\rm FIR}$ conversion factor reaches 0.4 and 0.8\,dex, and the differences in the SFR.vs.$L_{i, {\rm UV}}$ conversion factors reach 0.3 and 0.2\,dex when using S55 and NON-S55 IMF (including Cha03, K01, K93' and MS79 IMFs, partly caused by the difference in the IMF), respectively. At last, we give the conversion coefficients between SFR and these diagnostics for all models.
\end{abstract}


\section{Introduction}
One of the most recognizable features of galaxies along the Hubble sequence (loose definition, including not only morphological type but also gas content, mass, bar structure and dynamical environment) is the wide range in young stellar content and star formation activity. Understanding its physical nature and origin of the variation in stellar content are fundamental to understand evolution of galaxies \citep[][hereafter K98]{ken98}.
Star formation rate (SFR) can be used to compare with those distant galaxies at cosmological lookback times, and extrapolate the future timescales for star formation in galaxies by combining with HI and CO measurements \citep{ken94}.
Moreover, star formation activity is usually correlated with cold gas and stars in galaxies: stars continuously produce mass, energy and metals during their evolution processes, and return them to galactic medium (gas), affecting the status of the next generation of stars. SFR carries the information on the evolution of galaxies. Therefore, it is important to determine SFR and its variation with Hubble type (loose definition) and environment, which can help us to understand the evolution of galaxies.

The commonly used SFR tracers include the flux of H$\alpha$ nebular recombination line, the ultraviolet (UV) continuum flux, the flux of [OII]$\lambda$\,3727 forbidden-line doublet and far infra-red (FIR) continuum flux (K98). These SFR tracers are more or less correlated with the UV passband.

\begin{itemize}
\item First, the UV flux is directly tied to the photospheric emission of the young stellar population (SP).

\item Second, the integrated luminosity of galaxy shortward of the Lyman limit (Far-UV) can ionize the hydrogen in the nebula and produce the recombination lines (such as H$\alpha$, H$\beta$, and so on). Thus the luminosities of these lines can be used to trace SFR.

\item Third, the luminosity of the strong [OII]$\lambda$\,3727 forbidden-line doublet is often empirically obtained through the H$\alpha$ luminosity, although it is not coupled to the ionizing luminosity and the excitation of this line is sensitive to abundance and the ionization state of the gas.

\item Finally, the last SFR trace, the FIR luminosity, is also correlated with the UV passband. The interstellar dust can absorb the bolometric luminosity of galaxy and re-emit it in the thermal IR passband. The absorption cross section of the dust peaks in the UV passband, and, since the UV flux is considered as a tracer of young SP, the FIR luminosity can also diagnose SFR.
\end{itemize}

Furthermore, the last three diagnostics and the corresponding calibrations of SFR are correlated with the UV flux. Since
in our previous studies, we found that the inclusion of binary interactions in evolutionary population synthesis (EPS) models can raise the UV flux by $\sim$ 2-3 magnitudes for SP at an age of $\sim 1$\,Gyr \citep{zha04,zha05}, in this study we will discuss the effect of binary interactions on these calibrations.

The outline of the paper is as follows. In Section 2 we describe the used EPS models and algorithm. In Section 3 we overview the previous results about SFR calibrations, and the advantages and disadvantages of these SFR tracers. In Section 4 we give the effect of binary interactions on these SFR calibrations. In Section 5 we discuss the effects of initial mass function (IMF), gas-recycle assumption and the EPS models on these SFR calibrations, and give the conversion coefficients between SFR and these tracers for all models. Finally we present a summary and conclusions in Section 6.

\section{Models and algorithm}
\begin{table*}
\centering
\caption{Description of the used {\hf Yunnan, BC03, GISSEL98, PopSTAR, STARBURST99-v6.02} and {\hf P\'{E}GASE-v2} EPS models. The superscript 'C' means that it is available to choose, 'A' means that it is added by this work.}
\begin{tabular}{lrllrl}
\hline
 name      &  evolutionary LIB., $Z$, age range (yr)  & spectral LIB.  & IMF  & $M_{\rm l}, M_{\rm u}$    &OUTPUT  \\
           &                               &                &      & (M$_\odot)$ &  ISED/$Q$/$L_{\rm lines}$      \\
 \hline
  {\hf Yunnan}   & {\hf Cambridge}  ,solar, 0.1M-15G& {\hf BaSeL-2.0}  & S55/MS79$\rm ^{*1}$   & $\sim0.1$-100 &  Y/Y/Y   \\
  {\hf BC03}     & {\hf Padova}$\rm ^C$     ,solar, 1.0M-20G& {\hf BaSeL-3.1}$\rm ^C$  & S55/Cha03       & 0.10-100     & Y/N/N  \\
  {\hf GISSEL98} & {\hf Geneva}$\rm ^{*2}$     ,solar, 1.0M-20G& combination$\rm ^{*3}$      & MS79            &  0.10-100      & Y/N/N   \\
  {\hf PopSTAR}  & {\hf Padova}     ,solar, 0.1M-6.\,G& {\hf BaSeL}$\rm ^{*4}$  & S55'/K01$\rm ^C$ & 0.15-100      & N/Y/Y$\rm ^{*5}$\\
  {\hf STARBURST99-v6.02}& {\hf Padova} AGB$\rm ^C$ ,solar,  1.0M-15G& {\hf BaSeL}$\rm ^C$ & S55/K93'$\rm ^A$  & 0.10-100      & Y/Y/Y$\rm ^{*6}$\\
  {\hf P\'{E}GASE-v2}& {\hf Padova},  1.0M-20G    & {\hf BaSeL-2.0}+CM & S55/K93'$\rm ^C$ &  0.10-100     & Y/Y/Y$\rm ^{*7}$ \\
\hline
\end{tabular}
\begin{flushleft}
$*1$ In the {\hf Yunnan} models the IMF is for the primary in binary a system. \\
$*2$ {\hf GISSEL98} models use several stellar evolutionary libraries, {\hf Geneva} is the main one. \\
$*3$ {\hf GISSEL98} models use the combination of several stellar spectral libraries. \\
$*4$ {\hf PopSTAR} models use several stellar spectral libraries, {\hf BaSeL} is the main one. \\
$*5$ {\hf PopSTAR} models provide the number of ionizing photons $Q{\rm (H,HeI,HeII,OI)}$, the luminosities of emission-lines $L_{\rm (H\alpha, H\beta)}$. In the 2nd of a series, the luminosities of the other 18 emission-lines are provided as a function of $Q$(H) for HII region. \\
$*6$ {\hf STARBURST99} models provide $Q{\rm (H,HeI,HeII)}$ and $L_{(\rm H\alpha, H\beta, P\beta, BG)}$. \\
$*7$ {\hf P\'EGASE} models provide $Q{\rm (H)}$ and the luminosities of 61 spectral lines. \\
\end{flushleft}
\label{Tab:epsmod}
\end{table*}

\subsection{Spectral synthesis models}
\label{Sect:mod1}
First, we use the {\hf Yunnan} EPS models, which have been built by Zhang and her colleagues since 2002 \citep{zha02}. The main characteristic of {\hf Yunnan} EPS models is the inclusion of various binary interactions. {\hf Yunnan} EPS models have given the results of SPs with and without binary interactions at seven metallicities (from $0.0001$ to $0.03$) and 90 ages (from 0.1Myr to 15Gyr).

The {\hf Yunnan} EPS models were built on the basis of the {\hf Cambridge} stellar evolutionary tracks \citep{egg71,egg72,egg73}, {\hf BaSeL-2.0} stellar atmosphere models \citep[hereafter LCB97, LCB98]{lej97,lej98} and various initial distributions of stars. The {\hf Cambridge} stellar evolutionary tracks are obtained by using the rapid stellar evolution code \citep{hur00,hur02}, which is based on the stellar evolutionary tracks by \citet{pol98}.

In this work we use a set of standard {\hf Yunnan} EPS models at solar metallicity. The description of standard models is as follows: (i) the initial mass of the primary $M_1$ is given by
\begin{equation}
M_1 = \frac{0.19X}{(1-X)^{0.75} + 0.032 (1-X)^{0.25}},
\label{eq.imfms79-app}
\end{equation}
where $X$ is a random variable uniformly distributed in the range [0, 1]. This expression is the approximation by \citet[][hereafter EFT]{egg89} to the IMF ($\phi(M)= {\rm d}N / {\rm d}M
$) of \citet[][hereafter MS79]{mil79}:
\begin{equation}
\phi(M)_{_{\rm MS79}} \propto \Biggl\{ \matrix{
          M^{-1.4}, & 0.10 \le M \le 1.00 \cr
          M^{-2.5}, & 1.00 \le M \le 10.0 \cr
          M^{-3.3}, & 10.0 \le M \le 100. \cr
          }
\label{eq.imfms79}
\end{equation}
in which $M$ is the stellar mass in units of M$_{\rm \odot}$;
(ii) the initial masses of the component stars in a binary system are assumed to be correlated, and the initial mass of the secondary is obtained from a uniform mass-ratio $q$ distribution \citep{maz92,gol94};
(iii) the distribution of separation between two component stars takes the following form:
\begin{equation}
a \cdot {\rm n} (a)={\hf \bigl\{ \matrix{a_{\rm sep} (a/a_0)^m, & a \le
a_0 \hfill \cr a_{\rm sep}, \hfill & a_0 < a \le a_1 \cr}}
\label{eq.disa}
\end{equation}
in which $a_{\rm sep} \approx 0.070, a_0 = 10 {\rm R_\odot}, a_1 = 5.75 \times 10^6 {\rm R_\odot}$ and $m \approx 1.2$ \citep{han95};
and (iv) the eccentricity values follow a uniform distribution.

In the standard {\hf Yunnan} EPS models, approximately 50\% of stellar systems are binary systems with orbital periods less than 100yr. This fraction is a typical value for the Galaxy, resulting in $\sim10.1\%$ of the binaries experiencing Roche lobe overflow during the past 13\,Gyr (see Han et al. 1995).

To investigate the effect of IMF on the results, we use another set of solar-metallicity {\hf Yunnan} EPS models, which differs from the standard models by the initial mass distributions of the primary and secondary stars. In this set of models, (i) the initial mass of the primary is given by
\begin{equation}
M_1 = 0.3(\frac{X}{1-X})^{0.55}.
\label{eq.imfs55-app}
\end{equation}
This expression is the approximation by EFT to the IMF of \citet[][hereafter S55]{sal55} with $\alpha=-2.35$, i.e.,
\begin{equation}
\phi(M)_{\rm S55} = M^{-2.35};
\label{eq.imfs55}
\end{equation}
and (ii) the masses of the component stars in a binary system are assumed to be uncorrelated, i.e., the secondary mass is chosen independently from the same IMF as the primary.

In Table.~\ref{Tab:epsmod} we give the main characteristics of the used EPS models. The first column gives the name of each EPS model, the 2nd-5th columns give the used stellar evolutionary tracks (including metallicity $Z$ and the age range), the used stellar atmosphere models, the used IMF and the lower and upper mass limits ($M_{\rm l}$ and $M_{\rm u}$), and the last column describes whether the integrated spectral energy distributions (ISEDs), the number of ionizing photons $Q$ and the luminosities of spectral lines $L_{\rm lines}$ are provided by the corresponding models.

\subsubsection{Other spectral synthesis models}
\label{Sect:mod2}
To check the effect of spectral synthesis models on these SFR calibrations, we also use the {\hf GISSEL98} (Galaxy Isochrone Synthesis Spectral Evolution Library, \citealt{bru93}), {\hf BC03} \citep{bru03}, {\hf PopSTAR} \citep{mol09}, {\hf STARBURST99} v6.02 \citep{lei99,vaz05,lei10} and {\hf P\'{E}GASE} v2 \citep{fio97,fio99} models.
All these models do not include binary interactions, and the characteristics of them are also summarized in Table~\ref{Tab:epsmod}. Next we describe in more detail the main ingredients of these additional EPS models.

\vskip 0.4cm \leftline{\bf a. {\hf GISSEL98} and {\hf BC03} models}
{\hf GISSEL98} and {\hf BC03} models were built by Bruzual \& Charlot in 1993 and 2003, respectively, and provided the results of SPs at six metallicities (from 0.0001 to 0.05) and 221 ages (from 1\,Myr to 20\,Gyr) in tabular form.

{\hf GISSEL98} models were based on the {\hf Geneva} stellar evolutionary tracks \citep{mae89,mae91} mainly, the MS79 IMF with the lower and upper mass limits $M_{\rm l}=0.1$ and $M_{\rm u}=100.\rm M_\odot$ (see Eq.~\ref{eq.imfms79}), and the combination of several stellar spectral libraries (we refer the interested reader to their paper for details).

For the {\hf BC03} models, they used two sets of stellar evolutionary tracks ({\hf Padova}-1994 \& {\hf Padova}-2000), two forms of IMF [S55 with $\alpha = -2.35$ \& Chabrier (2003, hereafter Cha03)], high- [{\hf STELIB} \citep{LeB03} \& {\hf Pickles} \citep{pic98}] and low-resolution [{\hf BaSeL-3.1}, LCB97, LCB98] stellar spectral libraries. The Cha03 IMF is as follows:
\begin{equation}
\phi(M)_{_{\rm Cha03}} = \Bigl\{ \matrix {
          {\rm C_1} M^{-1} {\rm exp}^ {[{-({\rm log}M-{\rm logM_c})^2 \over 2\sigma^2}]}, & M \le 1.0\cr
          {\rm C_2} M^{-2.3} \hfill,  & M > 1.0\cr
          }
\label{eq.imfcha03}
\end{equation}
in which ${\rm M_c}=0.08$\,M$_\odot$, $\sigma=0.69$ and $M$ is the stellar mass in units of M$_\odot$. For both IMFs, the lower and upper mass limits are 0.1 and
100.$\rm M_\odot$.

For the {\hf GISSEL98} models, we use the set of ISEDs for solar-metallicity SPs. For the {\hf BC03} models we use the set of ISEDs of solar-metallicity SPs, which are derived by using the {\hf Padova}-1994 stellar evolutionary tracks and {\hf BaSeL-3.1} stellar spectral library.

\vskip 0.4cm \leftline{\bf b. {\hf PopSTAR} models}
The {\hf PopSTAR} models were built by \citet{mol09}. In their models, they used a revision of the {\hf Padova} \citep{bre98} isochrones used in \citet{gar98}, and the {\hf BaSeL} stellar atmosphere models (LCB97, LCB98) mainly.
All the results [the number of ionizing photons $Q{\rm (H, HeI, HeII, OI)}$ and the emission-line luminosities $L_{(\rm H\alpha, H\beta)}$] of SPs are provided in tabular form, including six metallicities, six IMFs and 93 ages (from 0.1Myr to log$t$/yr =9.78). We choose the two solar-metallicity sets, corresponding to the following IMFs:
\begin{itemize}
\item the S55 IMF with $\alpha=-2.35$ and mass range of $0.15-100\,\rm M_\odot$ (because the $M_{\rm l}$ is different from that of the other models, we call S55' IMF in Table~\ref{Tab:epsmod});
\item the Kroupa IMF (2001, hereafter K01)
\begin{equation}
\phi(M)_{_{\rm K01}} = \Biggl\{ \matrix {
          {\rm C_1} M^{-0.30}, & 0.01 \le M \le 0.08 \cr
          {\rm C_2} M^{-1.30}, & 0.08 \le M \le 0.50 \cr
          {\rm C_3} M^{-2.30}, & 0.50 \le M \le 100. \cr
          }.
\label{eq.imfk01}
\end{equation}
in which $M$ is the stellar mass in units of M$_\odot$.
\end{itemize}

Moreover in the 2nd paper of a series \citep{mar10}, the luminosities of other 18 emission-lines are given.

\vskip 0.4cm \leftline{\bf c. {\hf STARBURST99}-v6.02 code}
{\hf STARBURST99}\footnote{http://www.stsci.edu/science/starburst99/} is a web based software and data package designed to model spectrophotometric and related properties of star-forming galaxies. It was developed by researchers in Space Telescope Science Institute. We use the 6.02 version. The description of the model input physics is given by \citet{lei99}, \citet{vaz05} and \citet{lei10}.
\begin{itemize}
\item
For the {\hf STARBURST99}-v6.02 code, four sets of stellar evolutionary tracks are provided, each set including 5 metallicities. We choose the set of solar-metallicity {\hf Padova} AGB tracks (including thermally pulsing AGB stars, the 44th tracks).
\item
Five sets of stellar atmosphere libraries are provided, we choose the {\hf BaSeL} stellar atmosphere models (LCB97, LCB98).
\item
By default, the Kroupa IMF with two mass intervals (the exponent $\alpha=[-1.3, -2.3]$, the mass boundary $M_{\rm {cut}}$ = +[0.1 ,0.5, 100.]\,$\rm M_{\odot}$) is used. To be consistent with those of the other spectral synthesis models, we change this IMF form to (i) the S55 IMF with $\alpha=-2.35$ and (ii) the Kroupa (1993, hereafter K93) IMF with three mass intervals:
\begin{equation}
\phi(M)_{_{\rm K93}} = \Biggl\{ \matrix {
          {\rm C_1} M^{-1.3}, \ \ 0.10 \le M \le 0.50 \cr
          {\rm C_2} M^{-2.3}, \ \ 0.50 \le M \le 1.00 \cr
          {\rm C_3} M^{-2.7}, \ \ 1.00 \le M \le 100. \cr
          }
\label{eq.imfk93}
\end{equation}
in which ${\rm C_1}=0.035$, ${\rm C_2}=0.019$, ${\rm C_3}=0.019$ and $M$ is the stellar mass in units of M$_{\odot}$. Because all coefficients in Eq.~\ref{eq.imfk93} are set to 1 in this study (see also {\hf P\'{E}GASE}), we call K93' IMF in Table~\ref{Tab:epsmod}.
\item Two cases of star formation (const, without) are provided to choose, we choose the case of fixed mass (i.e., without star formation).
\end{itemize}

At last, we obtain the ISEDs, the number of ionizing photons $Q({\rm H, HeI, HeII})$, the emission-line luminosities $L_{\rm (H\alpha, H\beta, P\beta, BG)}$ of solar-metallicity SPs at 83 ages in the range 1Myr-15Gyr at an interval of log$t$/yr = 0.05.

\vskip 0.4cm \leftline{\bf d. {\hf P\'{E}GASE}-v2 code} {\hf P\'{E}GASE}
\footnote{http://www2.iap.fr/users/fioc/PEGASE.html} is a code to compute the spectral evolution of galaxies, and was developed by \citet{fio97,fio99}.
The evolution of the stars, gas and metals is followed for a law of star formation and a stellar IMF. The stellar evolutionary tracks extend from the main sequence to the white dwarf stage. The emission of the gas in HII regions is also taken into account. We use the 2.0 version. The main improvement in version 2 is the use of evolutionary tracks of different metallicities (from $10^{-4}$ to 5\,${\rm Z_\odot}$). The effect of extinction by dust is also modeled using a radiative transfer code.

\begin{itemize}
\item For the {\hf P\'{E}GASE}-v2 code, a set of {\hf Padova} stellar evolutionary tracks (7 metallicities, $Z$=0.0001, 0.0004, 0.004, 0.008, 0.02, 0.05 and 0.1), the combination of {\hf BaSeL}-2.0 (LCB97, LCB98, for $T_{\rm eff} \le 50\,000$\,K) with \citet[][hereafter CM, for $T_{\rm eff} > 50\,000$\,K]{cle87} stellar spectral libraries are used.
\item Among the nine IMFs provided, we choose the S55 IMF with $\alpha=-2.35$ and K93 IMF. The upper and lower mass limits of both IMFs are set to 0.1 and 100\,M$_\odot$ (by default, $M_{\rm u}$=120.\,M$_\odot$), respectively.
Because the coefficients of K93 IMF (see Eq.~\ref{eq.imfk93}) are set to 1 in the {\hf P\'{E}GASE} code, we also call K93' IMF in Table~\ref{Tab:epsmod}.
\item Among the six forms of SFR provided, we choose the instantaneous burst, the const SFR and the exponentially decreasing form ($SFR = p_2 \cdot {\rm exp}(-\tau/p_1)/p_1$, $p_2$=1.0) with the timescales ($p_1$) of $1,2,3,5,15$ and 30\,Gyr. They are used to built Burst, Irr, E-Sd  types of galaxies, respectively.
\item For the other model input parameters, the default values are used. Metallicity (mass fraction) of the interstellar medium (ISM) at $t=0$ is zero, the fraction of close binary systems is 0.5, and the mass fraction of sub-stellar objects formed is 0.0. And the default evolution processes are used, the evolution of stellar metallicity is consistent, stellar wind, infall, galactic winds and global extinction are neglected and nebular emission is considered.
\end{itemize}

Using the above set of input parameters and physics, we obtain the number of ionizing photons $Q$(H) and the luminosity of recombination line $L_{\rm H\alpha}$ for Burst, E, S0, Sa-Sd and Irr galaxies at 68 ages in the range of 1Myr-20Gyr.

\subsubsection{Comments on the above mentioned models}
For {\hf P\'{E}GASE} models we directly use their results because they consider the star formation and nebular emission. For the other models ({\hf GISSEL98, BC03, PopSTAR and STARBURST99}), we need to generate the results of galaxies with different galaxy types by means of spectral synthesis models.
Because the ISEDs are not provided for {\hf PopSTAR} models, we could not give the ISEDs of galaxies with different galaxy types (therefore the UV and FIR continuum fluxes), we only could give the luminosities of the H$\alpha$ recombination line and the [OII]$\lambda$3727 forbidden-line doublet, and the nebular emission continuum by using the number of ionizing photons $Q$(H) provided by them.

\subsection{Construction of galaxies with different galaxy types}
\label{Sect:csp}
The construction of galaxies with different galaxy types has been described in our previous paper \citep{zha10}. In brief, we use the {\hf BC03} software package to build them. Eight galaxy types (Burst, E, S0, Sa-Sd and Irr) are included, and they are built by a delta-form SFR, six exponentially decreasing SFRs with characteristic time-decays $\tau =1, 2, 3, 5, 10, 15$, and 30\,Gyr and a constant-form SFR, respectively. The exponentially decreasing SFR is given by
\begin{equation}
\psi(t) = [1 + \epsilon M_{\rm PG} (t)] \tau ^{-1} {\rm exp}(-t/\tau),
\label{eq.sfr}
\end{equation}
where $\tau$ is the e-folding timescale, $M_{\rm PG} (t)$ = $[1-{\rm exp}(-t/\tau)] - M_{\rm stars} - M_{\rm remnants}$ is the mass of gas that has been processed into stars and then returned to the ISM at time $t$, $M_{\rm stars}$ and $M_{\rm remnants}$ are the masses of stars and remnants at $t$, and $\varepsilon$ denotes the fraction of $M_{\rm {PG}} (t)$ that can be recycled into new star formation.

\subsection{Computations of $L_{\rm H\alpha}$, $L_{\rm [OII]}$, $L_{\rm FIR}$ and nebular continuum}
\label{Sect:cal}
In this part, we will describe the computations of the luminosity of the H$\alpha$ recombination line $L_{\rm H\alpha}$, the luminosity of the [OII]$\lambda$3727 forbidden-line double $L_{\rm [OII]}$, the FIR flux $L_{\rm FIR}$ and the emission of nebular continuum $F_{\rm neb, \lambda}$.

Before the computation of $L_{\rm H\alpha}$, we must calculate the number of ionizing photons $Q$(H). During the computation of $Q$(H), one method assumes that it only comes from young ($<$ 10\,Myr) and massive ($\ga$20-25\,${\rm M_\odot}$) stars (such as K98, {\hf STARBURST99}). Another method obtains that number by integrating the photons below 912$\rm \AA$ (such as {\hf PopSTAR}). We choose the second method, i.e.,
\begin{equation}
Q(\rm H) = \int {F_{\rm \nu} \over h\nu} d\nu, \label{Eq.qh}
\end{equation}
where $F_{\rm \nu}$ is the stellar flux in Hz, $\nu$ is frequency and h is the Planck constant (=\,$6.6262 \times 10^{-27}$\,erg\,s).

Then, we assume that all the star formation is traced by the ionized gas, although in the {\hf P\'{E}GASE} code 70 per cent of the Lyman continuum photons computed from the spectral synthesis models are absorbed by the gas (i.e., ionize the gas) and the rest (30\%) are absorbed by dust when considering extinction.
At last, we use Case B recombination at electron temperature $T_e$=10,000K and number density $n_e=100{\rm cm^{-3}}$. The same set of the parameter values (Case B, $T_e$ and $n_e$) is used by K98, {\hf STARUBURST99} and other studies.
Under the above assumptions, the luminosity of H$\alpha$ can be obtained by the following expression ({\hf PopSTAR}):
\begin{equation}
L_{\rm H\alpha} = Q({\rm H}) {\alpha \over \beta} {j_{\rm B} \over
{\alpha_{\rm B}}},
\label{Eq.lha}
\end{equation}
where $\alpha_{\rm B}$ is the recombination coefficient to the excited level in hydrogen, which depends on the electronic temperature, $j_{\rm B}$ and $\alpha_{\rm B}$ are from \citet{fer80}, and the ratio ${\alpha / \beta}$ is taken from \citet{ost89}.

The luminosity of [OII]$\lambda$3727, $L_{\rm [OII]}$, is often obtained by an empirical method. In this paper we obtain it by assuming the ratio of the luminosity of [OII]$\lambda$3727$\rm \AA$ to H$\alpha$, $L_{\rm [OII]}/L_{\rm H\alpha} =0.45$, as used by K98, although in the {\hf P\'EGASE} code it equals to 3.01/2.915, and the discrepancy in the $L_{\rm [OII]}/L_{\rm H\alpha}$ is large for the different types of galaxies.

The FIR luminosity is mainly used to calibrate the SFR of starburst galaxies, for which the FIR luminosity is often assumed to equal to the bolometric luminosity, $L_{\rm FIR} = L_{\rm BOL}$.

At last, the emission of nebular continuum can be obtained by:
\begin{equation}
F_{\rm neb, \lambda} = {\Gamma {c \over \lambda^2 \alpha_{\rm B}} Q(\rm
H)}, \label{Eq.fneb}
\end{equation}
where $c$ is the light velocity and $\Gamma$ is the emission coefficient for hydrogen and helium (He/H=0.1), which includes free-free and free-bound contributions and the emission coefficient due to the two-photon continuum. The $\Gamma$ coefficient is wavelength-dependent and is taken from \citet{all84} and \citet{fer80}.

\section{Overview of SFR calibrations and properties of SFR tracers}
\label{Sect:rst-cp}
In order to compare our derived results with the previous studies, in this part we will overview the previous results about SFR calibrations in terms of luminosities of recombination-line, forbidden-line, UV and FIR continuum, and overview the properties (advantages and drawbacks) of these SFR tracers.

\subsection{Overview of $SFR$.vs.$L_{\rm UV}$}
The UV luminosity, $L_{\rm UV}$, is directly tied to the emission of the young SP. For convenience, it is often expressed as the linear relation with SFR by previous studies: $SFR_{\rm UV}={\rm F}_{\rm UV} \times L_{\rm UV}$, where $SFR_{\rm UV}$ and $L_{\rm UV}$ are in units of $\rm M_\odot\,yr^{-1}$ and erg\,s$^{-1}$\,Hz$^{-1}$.
This kind of SFR calibration is valid only for galaxies with continuous SFR over timescales of 10$^8$ or longer (K98). Its advantage is that it can be applied to star-forming galaxies over a wide range of redshifts. Its drawback is that it is sensitive to extinction and the form of the IMF.

Because different EPS models and methods are used, the obtained conversion factor $\rm F_{\rm UV}$ between $SFR_{\rm UV}$ and $L_{\rm UV}$ is different, being the difference as big as 0.3\,dex (K98):
\begin{itemize}
\item K98 has obtained ${\rm F_{UV}} = 1.4 \times 10^{-28}$ by using the S55 IMF with mass limits 0.1 and 100\,M$_\odot$ and solar abundance.
\item \citet[][hereafter MPD98]{mad98} have obtained ${\rm F_{UV}} = (1.25 \times 10^{-28}, 1.26 \times 10^{-28})$ at (1500\,$\rm \AA$, 2800\,$\rm \AA$) by using the S55 IMF, and ${\rm F_{UV}}$ = $(2.86\times10^{-28}, 1.96\times10^{-28})$ by using the \citet{sca86} IMF. In their studies the calibration factors are from models with an exponentially decreasing SFR.
\item \citet[hereafter G10]{gil10} have obtained ${\rm F_{UV}} = 0.71 \times 10^{-28}$ by using the {\hf P\'{E}GASE} code (assuming a constant SFR over 1\,Gyr and a Kroupa IMF around solar metallicity).
\end{itemize}

\subsection{Overview of $SFR$.vs.$L_{\rm H\alpha}$}
Also, the luminosity of H$\alpha$ recombination line, $L_{\rm H\alpha}$, is often expressed as the linear relation with SFR: $SFR_{\rm H\alpha} = {\rm F}_{\rm H\alpha} \times L_{\rm H\alpha}$, where $SFR_{\rm H\alpha}$ and $L_{\rm H\alpha}$ are in units of $\rm M_\odot\,yr^{-1}$ and erg\,s$^{-1}$. This equation traces the instantaneous SFR.
Its primary advantage is its high sensitivity. In addition, H$\alpha$ is typically so conspicuous that it can easily be detected. Its drawback is that it is sensitive to extinction, IMF and the assumption that all of the massive star formation is traced by the ionizing gas.
The conversion factor ${\rm F}_{\rm H\alpha}$ obtained by different studies is as follows:
\begin{itemize}
\item K98 has obtained ${F_{H\alpha}} =7.9 \times 10^{-42}$ by using the S55 IMF with $\alpha=2.35$ and the upper and lower mass limits $M_{\rm l}=0.1$ and $M_{\rm u}=100.\, {\rm M}_\odot$. This value is obtained from a constant star formation model at 'equilibrium'. The similar value has also been obtained by \citet{ken94} and MPD98 when using the S55 IMF.
This factor has been adopted by \citet{shi06} and G10. In the work of G10, they have pointed out that the factor is similar to that obtained from the {\hf P\'{E}GASE} code.
\item \citet[][hereafter B04]{bri04} have obtained ${\rm {F_{H\alpha}}} =5.25 \times 10^{-42}$ by using K01 IMF. This value is multiplied by 1.5 in the works of  \citealt{tre04} and G10 in order to convert to the S55 IMF. A factor of 1.5 corresponds to the ratio of mass in the two IMFs for the same amount of ionizing radiation.
\item \citet{mat07} have obtained ${\rm {F_{H\alpha}}} = 5.22 \times 10^{-42} \times 10^{-0.4(r-r_{\rm fibre})}$, where $r$ is the \textit{Petrosian} magnitude representing the total galaxy flux, $r_{\rm fibre}$ is the $r-$band \textit{fibre} magnitude, and the last term is for aperture effect.
This equation is derived by using the expression adapted from K98 for a Cha03 IMF ($0.1-100\rm M_\odot$) and the prescriptions given by \citet[][hereafter H03]{hop03}.
This conversion factor has been used by \citet{hua09}.
\item K98 argued that when adopting the \citet{sca86} IMF, the derived $SFR_{\rm H\alpha}$ is approximately three times higher than that derived with a Salpeter IMF.
\end{itemize}

\subsection{Overview of $SFR$.vs.$L_{\rm [OII]}$}
The [OII]$\lambda$3727\,$\rm \AA$ forbidden-line doublet is one of the strongest emission lines in the blue. It is accessible to optical observations over a wide range of redshifts. Therefore, it provides a very useful estimate of the SFR in distant galaxies \citep[][K98]{sch99}.
Its disadvantage is less precise than that from H$\alpha$ and may also be prone to systematic errors from extinction and variations in the diffuse gas fraction.
Often, SFR is expressed as the linear relation with the luminosity $L_{\rm [OII]}$: $SFR_{\rm [OII]} = {\rm F}_{\rm [OII]} \times L_{\rm [OII]}$, where $SFR_{\rm [OII]}$ and $L_{\rm [OII]}$ are in units of $\rm M_\odot\,yr^{-1}$ and erg\,s$^{-1}$.
\begin{itemize}
\item K98 has obtained ${\rm F_{[OII]}} = (1.4 \pm 0.4) \times 10^{-41}$ by using $L_{\rm [OII]}/L_{\rm H\alpha}=0.45$.
\item H03 have obtained ${\rm F_{[OII]}} = 3.37 \times 10^{-41}$ by using $L_{\rm [OII]}/L_{\rm H\alpha}=0.23$ and the calibration factor ${\rm F_{H\alpha}} =7.9 \times 10^{-42}$ of K98.
\item G10 have obtained ${\rm F_{[OII]}} = 3.95 \times 10^{-41}$ by assuming the ratio of extinguished [OII] to H$\alpha$ flux $L_{\rm [OII]}/L_{\rm H\alpha} =0.5$, 1 mag of extinction at H$\alpha$ and the calibration factor ${\rm F_{H\alpha}} =7.9 \times 10^{-42}$ of K98. If neglecting the extinction, the G10 conversion factor ${\rm F_{[OII]}} = 1.57 \times 10^{-41}$.
\end{itemize}

\subsection{Overview of $SFR$.vs.$L_{FIR}$}
The FIR (10-300\,$\mu$m) luminosity, $L_{\rm FIR}$, is only used to trace SFR of starburst galaxies with ages less than $10^8$ years. Its drawback is that it is also sensitive to IMF and extinction.
Using the models of \citet{lei95} for continuous bursts of age 10-100\,Myr and the S55 IMF with the lower and upper mass limits $M_{\rm l}=0.1$ and $M_{\rm u} = 100\,\rm M_{\sun}$, K98 has presented the conversion factor between $SFR_{\rm FIR}$ and $L_{\rm FIR}$: $\rm F_{FIR} = 4.5 \times 10^{-44}$\,M$_\odot$\,yr$^{-1}$/erg\,s$^{-1}$.

\section{Effect of binary interactions on SFR calibrations and Comparisons}

\begin{figure}
\centering
\includegraphics[height=8.0cm,width=6.5cm,angle=270]{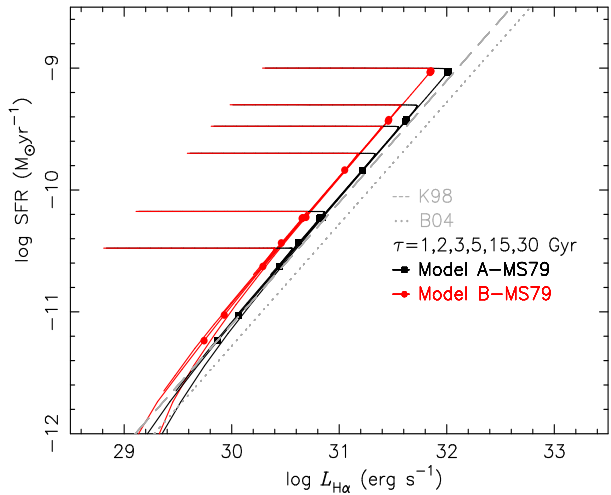}
\caption{Relation between $SFR$ and $L_{\rm H\alpha}$ of E, S0, Sa, Sb, Sc and Sd galaxies (corresponding to $\tau=1,2,3,5,15$ and 30\,Gyr in Eq.~\ref{eq.sfr}, from top to bottom) for Models A-MS79 (black solid line+solid rectangles) and B-MS79 (red solid line+solid circles). The ages of galaxies are in the range of 1\,Myr-15\,Gyr. Also shown are the results of K98 (dashed line) and B04 (dotted line).}
\label{Fig:sfr-lha}
\end{figure}

\begin{figure}
\centering
\includegraphics[height=8.0cm,width=6.5cm,angle=270]{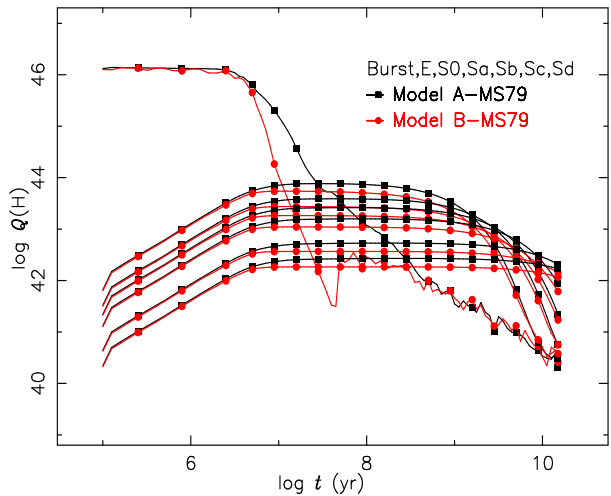}
\caption{The evolution of $Q$(H) for Models A-MS79 (black line+solid rectangles) and B-MS79 (red line+solid circles). From top to bottom are Burst, E, S0, Sa, Sb, Sc and Sd types of galaxies.}
\label{Fig:qh-t}
\end{figure}

\begin{figure}
\centering
\includegraphics[height=8.0cm,width=6.5cm,angle=270]{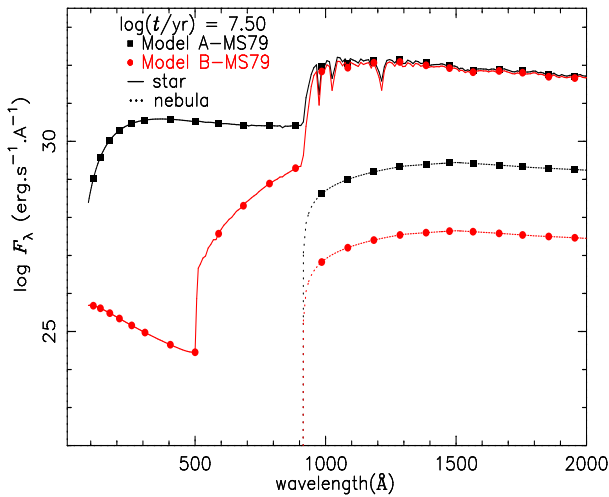}
\caption{The stellar (solid line) and nebular (dotted line) spectra of the Burst galaxy type (SP) at an age of log$t$/yr =7.5 for Models A-MS79 (black+rectangles) and  B-MS79 (red+circles).}
\label{Fig:ised-t36}
\end{figure}

\begin{figure*}
\includegraphics[angle=270,scale=.400]{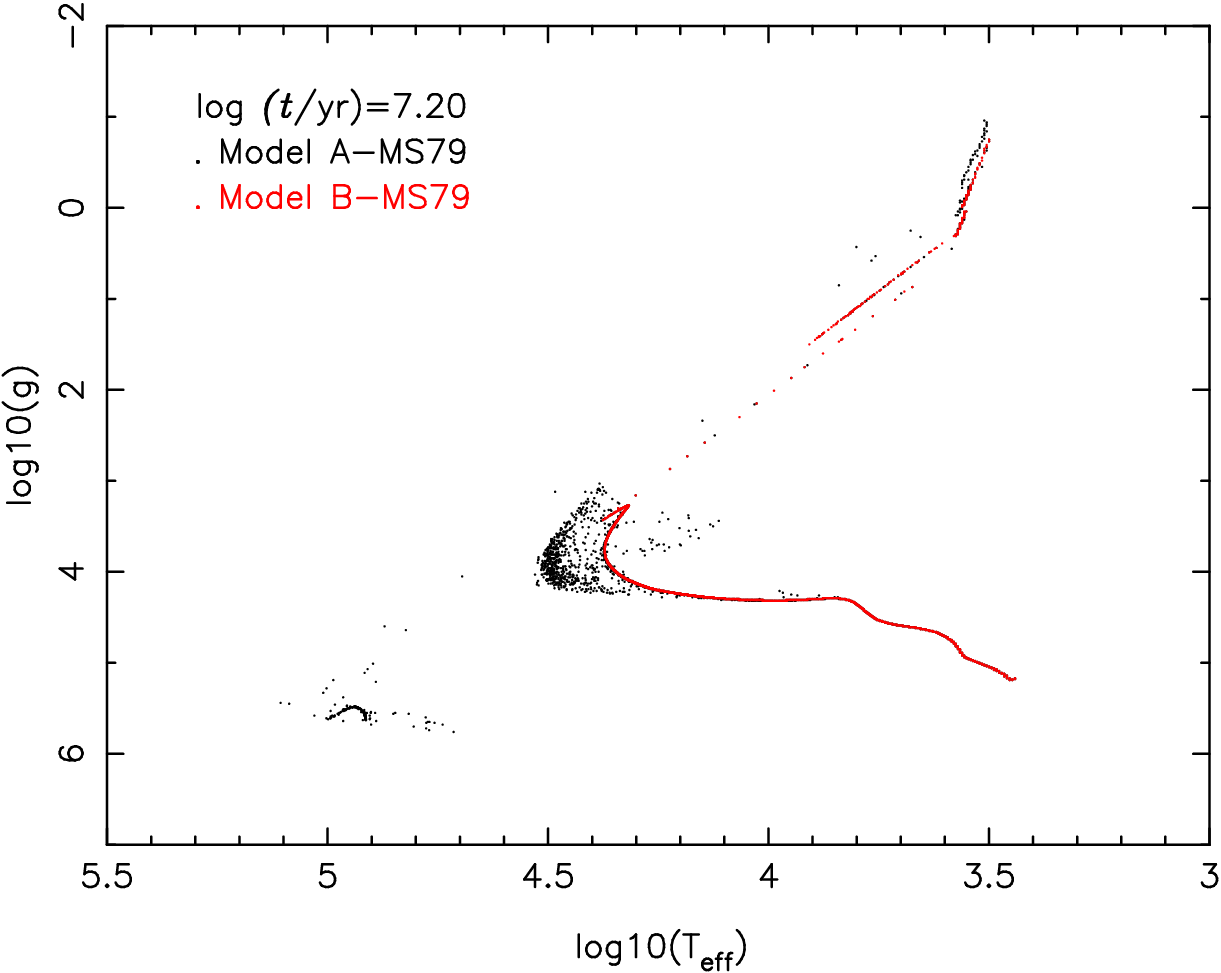}
\includegraphics[angle=270,scale=.400]{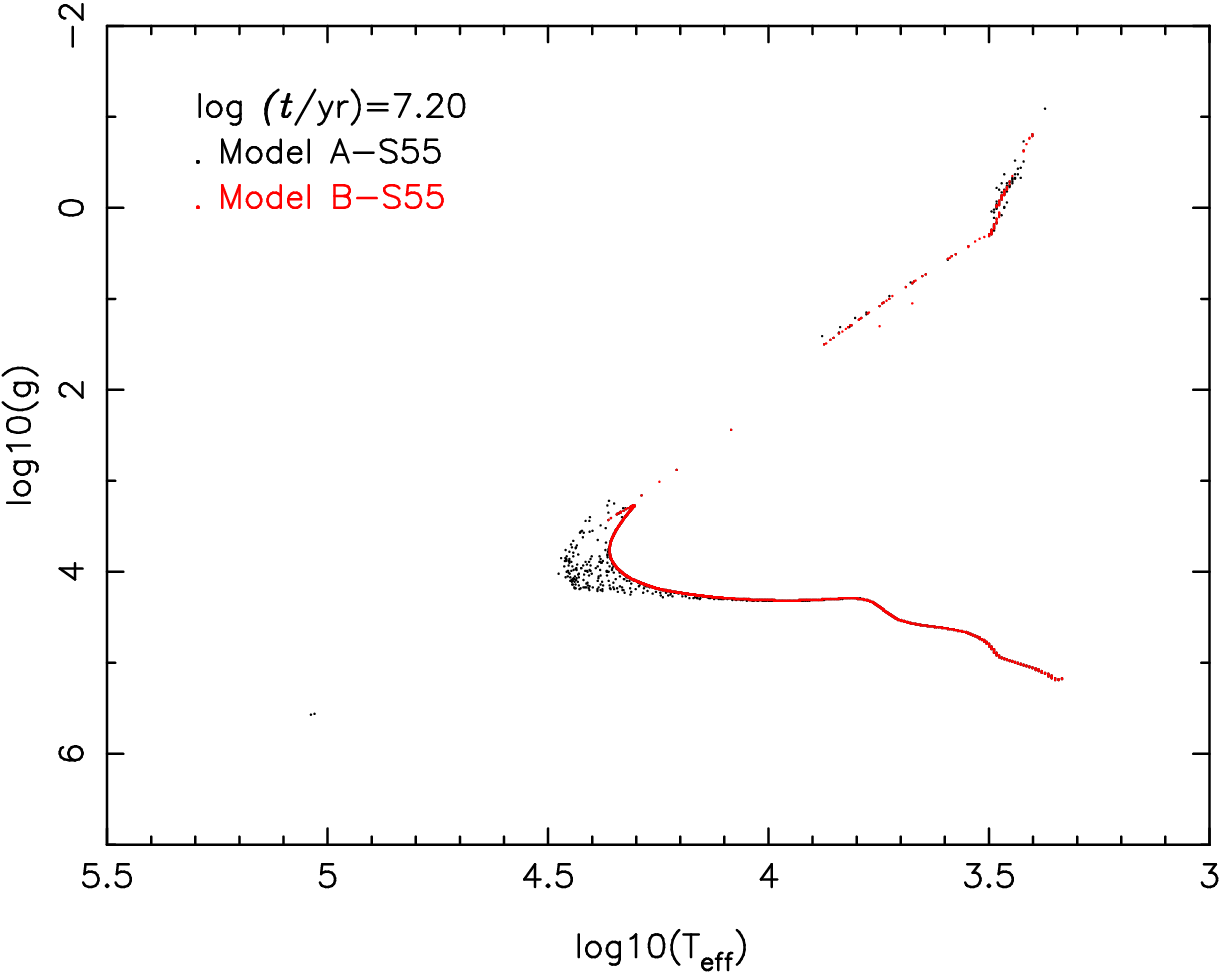}
\caption{The distribution of stars in the log$T_{\rm eff}$-log$g$ plane for the Burst galaxy type (SP) at an age of log$t$/yr =7.2. In each panel black and red symbols are for considering (i.e., Model A-X) and neglecting (i.e., Model B-X) binary interactions. Left and right panels are for Models A/B-MS79 and A/B-S55, respectively.}
\label{Fig:cmd}
\end{figure*}

\begin{figure}
\centering
\includegraphics[height=8.0cm,width=6.5cm,angle=270]{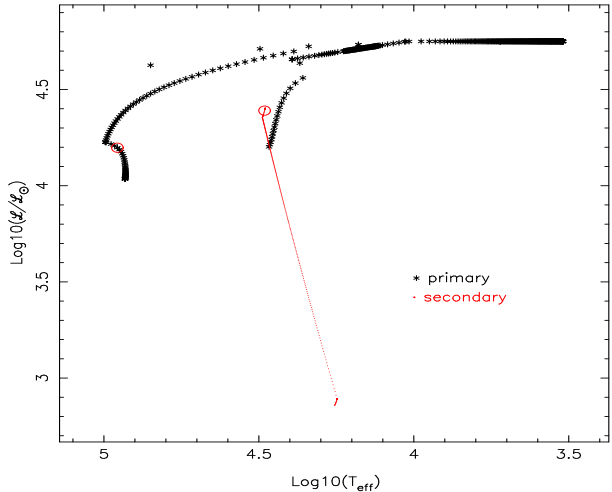}
\caption{The evolution of a binary system ($M_1=13.97, M_2=5.44$\,M$_\odot$ and $P=28.0$\,day) from ZAMS to the end of helium burning. Black and red symbols are for the primary and the secondary, respectively. The red open circles denote the positions at an age of log$t$/yr = 7.2, corresponding to time in Fig.~\ref{Fig:cmd}.}
\label{Fig:evo}
\end{figure}

\begin{figure*}
\includegraphics[bb=134 119 477 624,clip,height=8.0cm,width=4.5cm,angle=270]{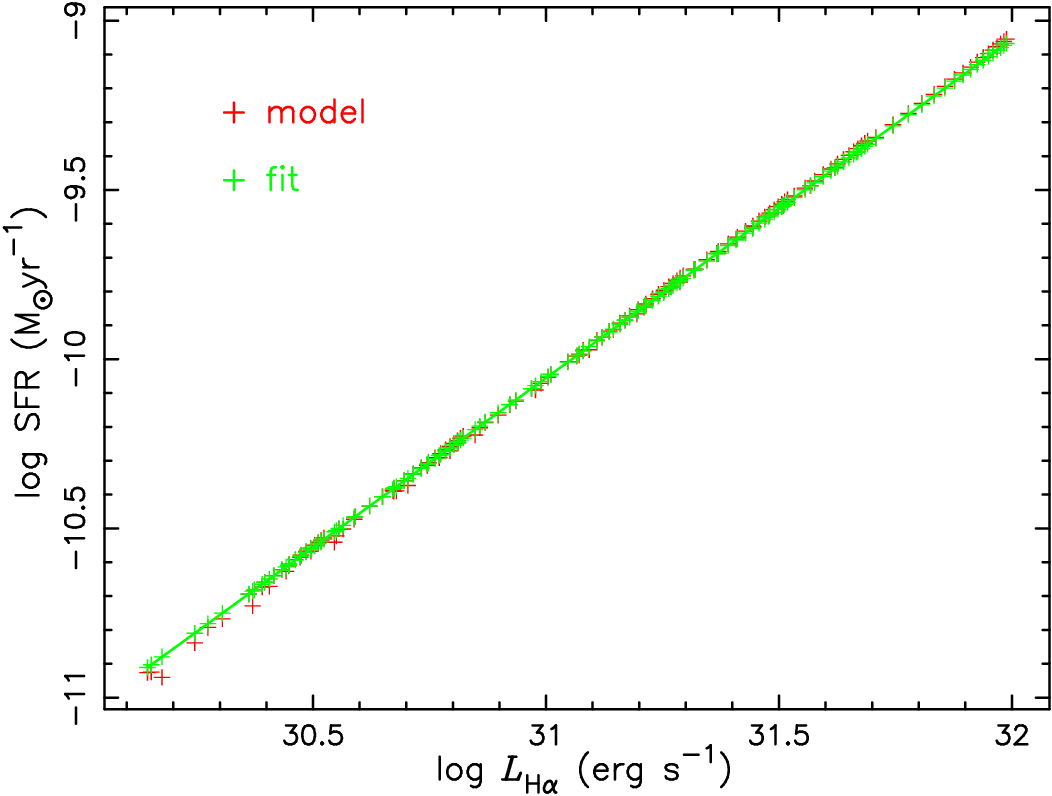}
\includegraphics[bb=134 119 477 624,clip,height=8.0cm,width=4.5cm,angle=270]{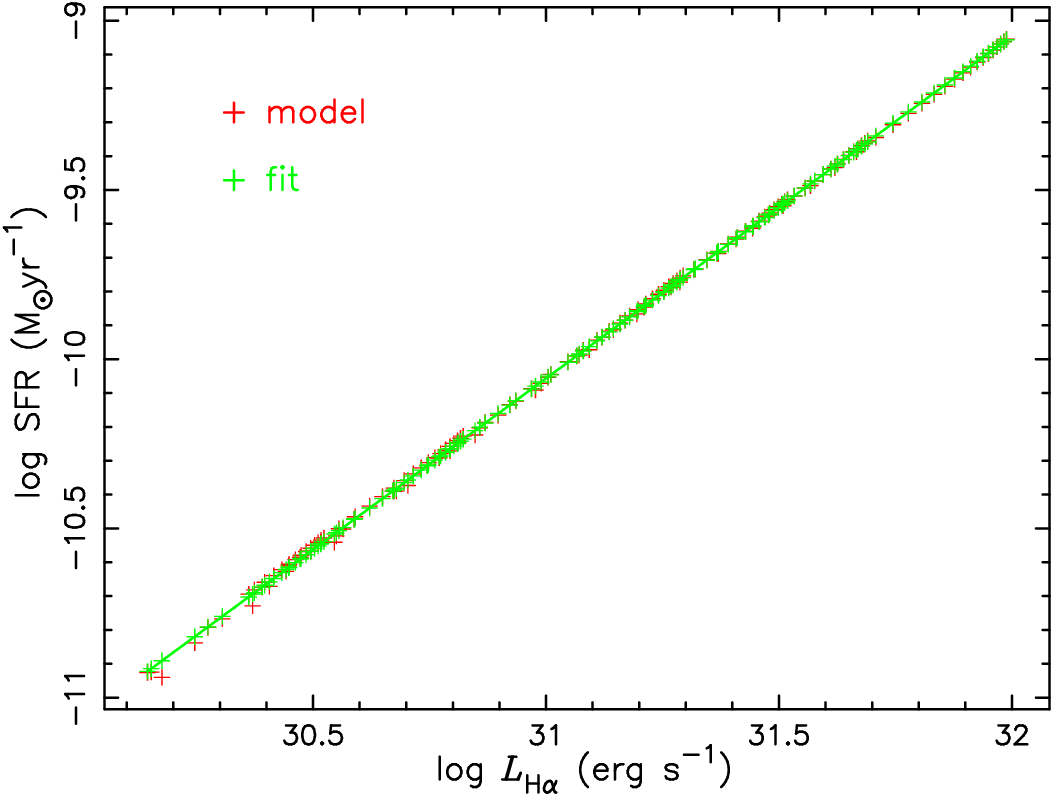}
\includegraphics[height=8.0cm,width=3.3cm,angle=270]{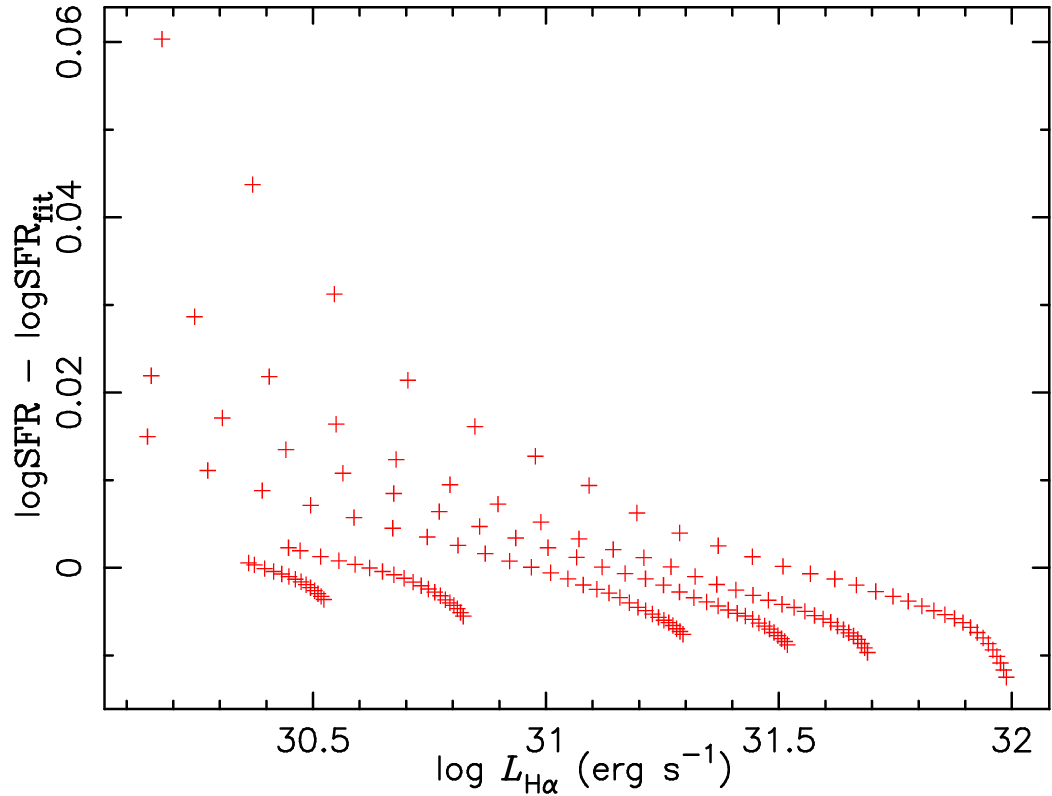}
\includegraphics[height=8.0cm,width=3.3cm,angle=270]{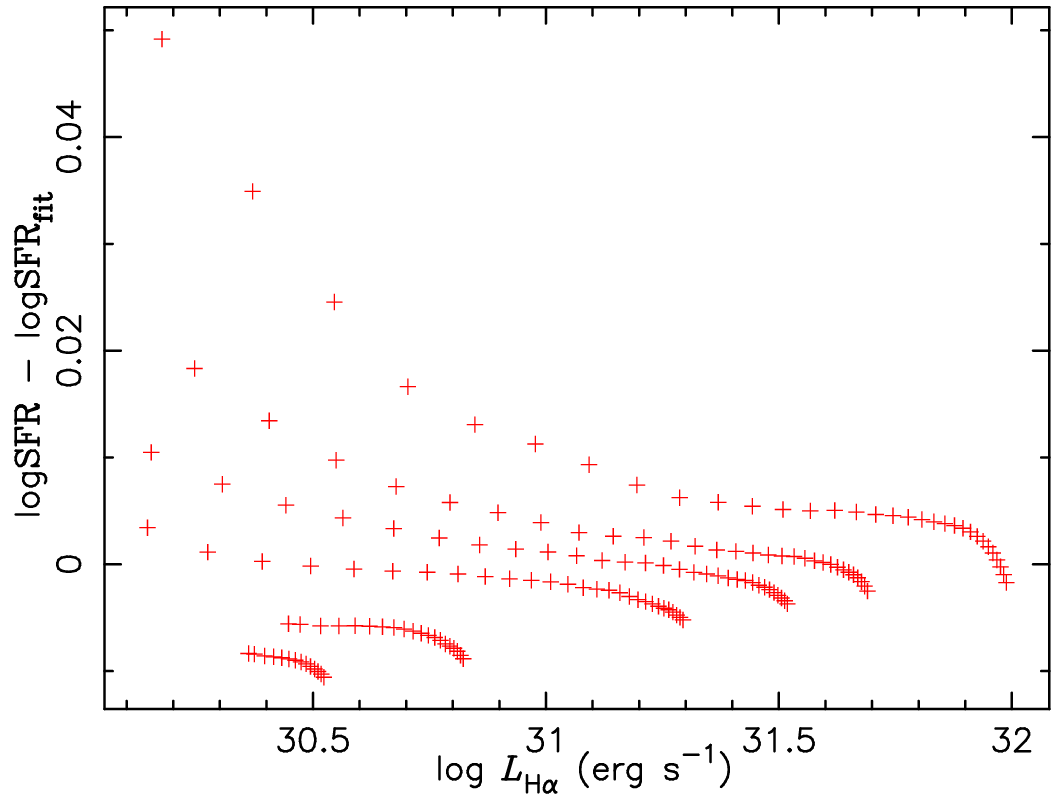}
\caption{Top panels are the comparison in $SFR$-$L_{\rm H\alpha}$ relation between model (red symbols) and fitting (green symbols) results for Model A-MS79. Left and right panels are for using Eqs.~\ref{Eq.ha-c1} and \ref{Eq.ha-cx}, respectively. Bottom panels are the residuals (log$SFR$-log$SFR_{\rm fit}$) as a function of log$L_{\rm H\alpha}$.}
\label{Fig:ha-fit}
\end{figure*}

\begin{figure*}
\includegraphics[angle=270,scale=.400]{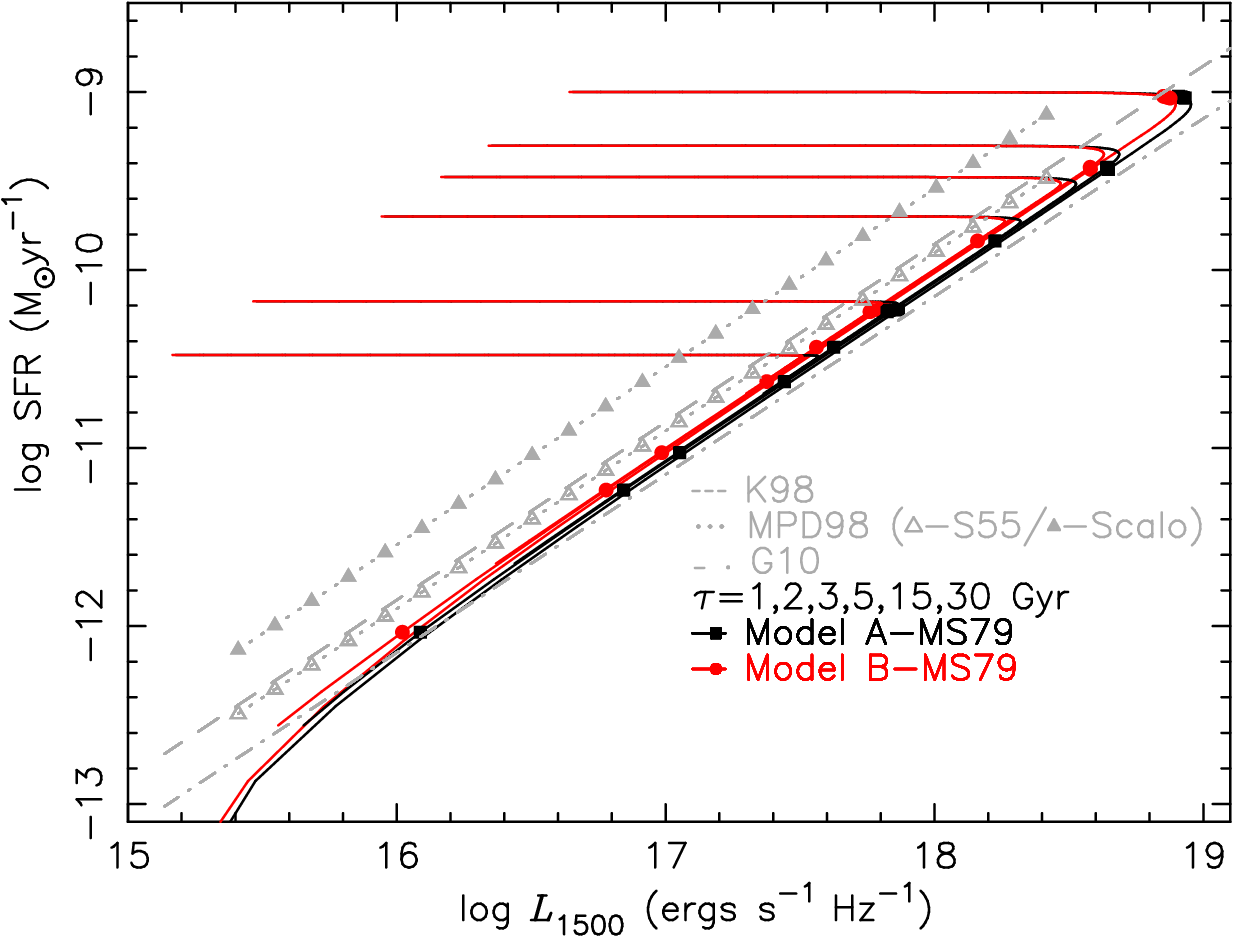}
\includegraphics[angle=270,scale=.400]{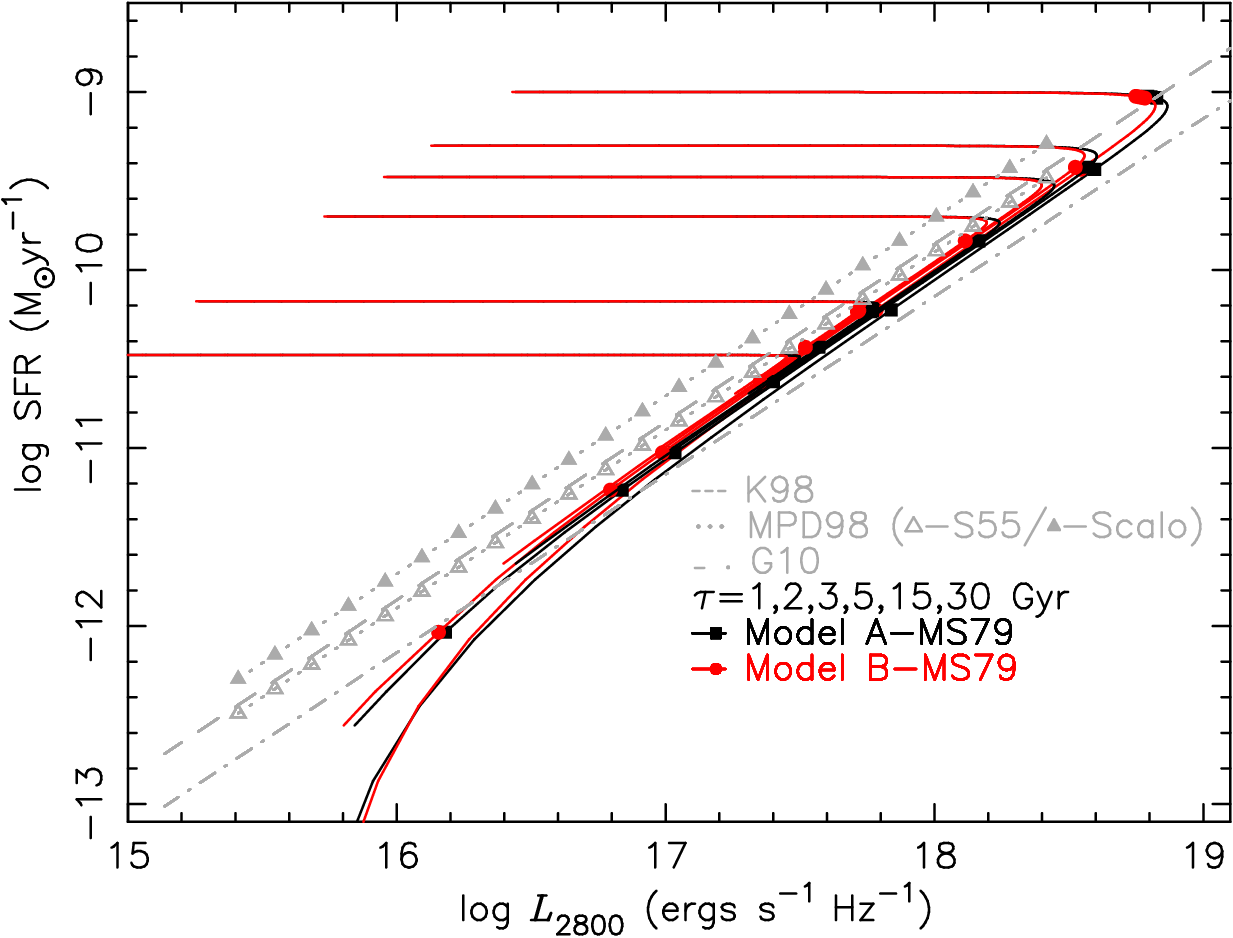}
\caption{Relations between $SFR$ and $L_{i,{\rm UV}}$ of E, S0, Sa, Sb, Sc and Sd galaxies (corresponding to $\tau=1,2,3,5,15$ and 30\,Gyr in Eq.~\ref{eq.sfr}, from top to bottom) for Models A-MS79 (black solid line+solid rectangles) and B-MS79 (red solid line+solid circles). Left panel is for $L_{\rm 1500}$, and right panel is for $L_{\rm 2800}$. Also shown are the results of K98 (grey dashed line), MPD98 (grey dotted line, open and solid triangles are for using the S55 and Scalo IMFs, respectively) and G10 (grey dash-dotted line).} \label{Fig:sfr-uv}
\end{figure*}

\begin{figure*}
\includegraphics[angle=270,scale=.400]{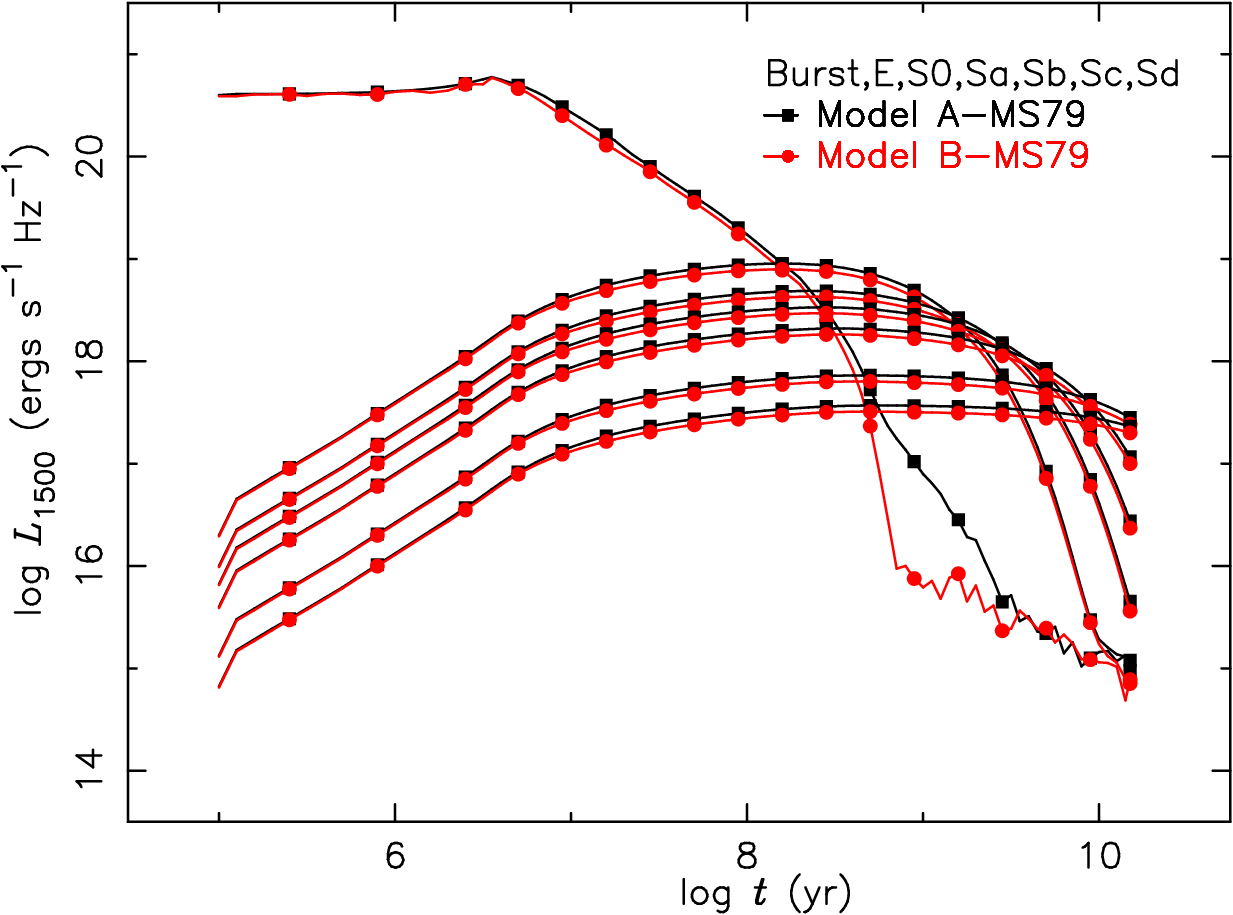}
\includegraphics[angle=270,scale=.400]{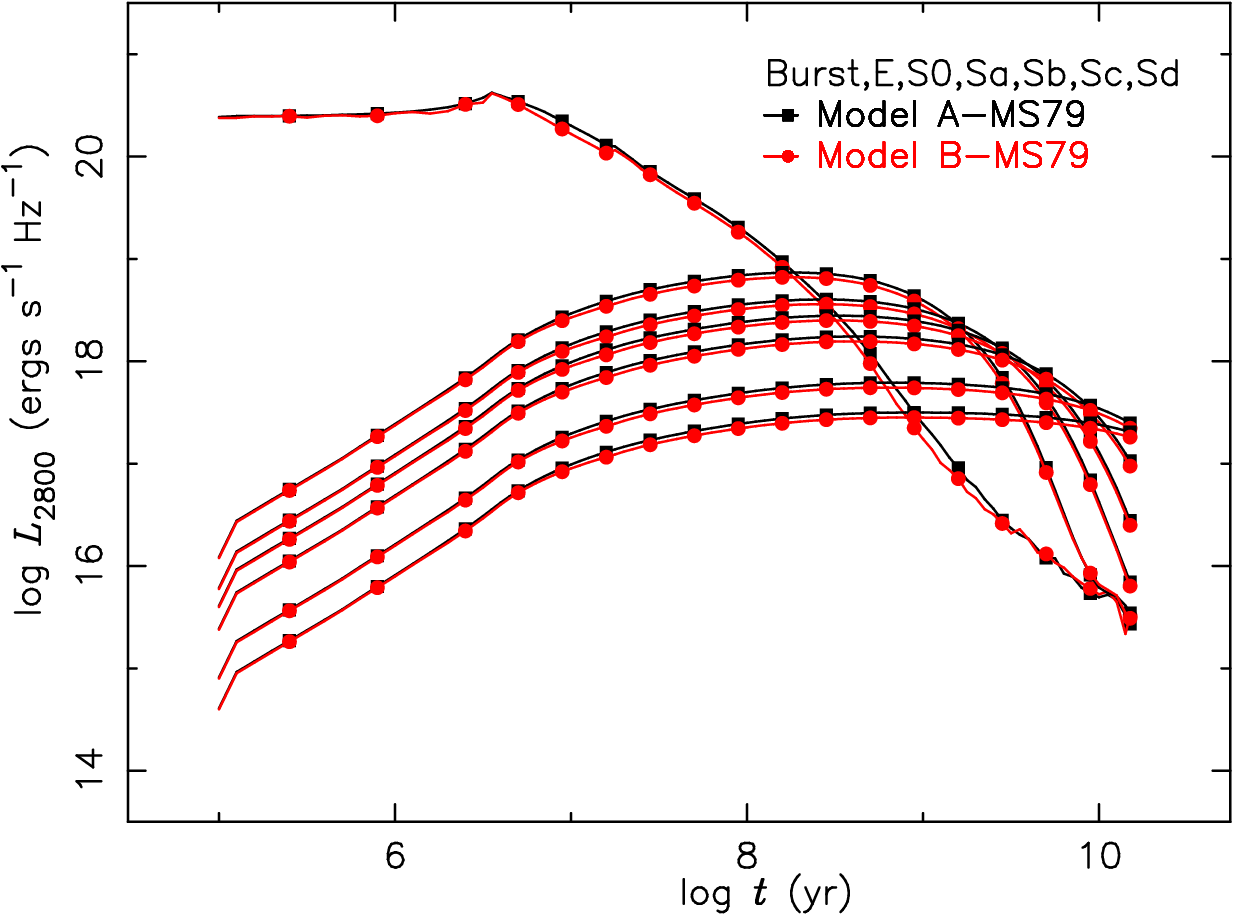}
\caption{Similar to Fig.~\ref{Fig:qh-t}, but the left panel is for $L_{1500}$ and the right panel is for $L_{2800}$.}
\label{Fig:uv-t}
\end{figure*}

\begin{figure}
\centering
\includegraphics[height=8.0cm,width=6.5cm,angle=270]{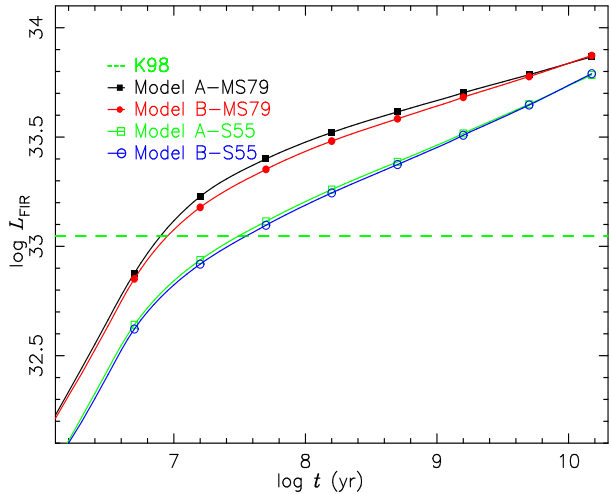}
\caption{The $L_{\rm FIR}$ evolution of Irr galaxies (i.e., models with constant star formation, SFR=1\,M$_{\rm \odot}$) for Models A-MS79 (black solid line+solid rectangles), B-MS79 (red solid line+solid circles), A-S55 (green solid line+open rectangles) and B-S55 (blue solid line+open circles). Also shown is the result of K98 (green dashed line).}
\label{Fig:fir-t}
\end{figure}

\begin{table}
\centering
\caption{Definitions of models. BIs and GR indicate binary interactions and gas recycle, respectively.}
\begin{tabular}{lrr lr}
\hline
  Model &  EPS              & BIs & IMF        & GR     \\
\hline
  A     & {\hf Yunnan}      & Y  & MS79/S55   & N       \\
  B     & {\hf Yunnan}      & N  & MS79/S55   & N       \\
  C     & {\hf BC03}        & N  & Cha03/S55  & N       \\
  D     & {\hf GISSEL98}    & N  & MS79       & N       \\
  E     & {\hf PopSTAR}     & N  & K01/S55'   & N       \\
  F     & {\hf STARBURST99} & N  & K93'/S55   & N       \\
  G     & {\hf P\'{E}GASE  }& N  & K93'/S55   & N       \\
  Cr    & {\hf BC03}        & N  & Cha03/S55  &{\bf Y}  \\
\hline
\end{tabular}
\label{Tab:defmod}
\end{table}

\begin{table}
\small
\centering
\caption{Conversion coefficients between $SFR$ and $L_{\rm H\alpha}$ for all models except for Models Cr-Cha03 and Cr-S55.}
\begin{tabular}{lrr lrr}
\hline
  Model & \multicolumn{2}{c}{Eq.~\ref{Eq.ha-c1}} & \multicolumn{3}{c}{Eq.~\ref{Eq.ha-cx}} \\
        & (C$_{\rm H{\alpha}}$ & $\sigma_{\rm H{\alpha}}$) & (C$_{\rm H{\alpha}}'$ & A$_{\rm H{\alpha}}'$ & $\sigma_{\rm H{\alpha}}'$)\\
\hline
A-MS79    &$  -41.0556$&$    0.0096$&$  -41.4316$&$    1.0121$&$    0.0076$\\
B-MS79    &$  -40.8942$&$    0.0141$&$  -41.4904$&$    1.0193$&$    0.0104$\\
A-S55     &$  -40.7408$&$    0.0206$&$  -41.6298$&$    1.0289$&$    0.0151$\\
B-S55     &$  -40.6939$&$    0.0216$&$  -41.6182$&$    1.0301$&$    0.0159$\\
C-Cha03   &$  -41.3025$&$    0.0054$&$  -41.5413$&$    1.0077$&$    0.0040$\\
C-S55     &$  -41.0785$&$    0.0065$&$  -41.3640$&$    1.0093$&$    0.0048$\\
D-MS79    &$  -41.2936$&$    0.0055$&$  -41.5374$&$    1.0079$&$    0.0041$\\
E-K01     &$  -40.7390$&$    0.0003$&$  -40.7296$&$    0.9997$&$    0.0002$\\
E-S55'    &$  -41.5047$&$    0.0002$&$  -41.4959$&$    0.9997$&$    0.0002$\\
F-K93'    &$  -40.4930$&$    0.0004$&$  -40.4773$&$    0.9995$&$    0.0004$\\
F-S55     &$  -40.8002$&$    0.0004$&$  -40.7850$&$    0.9995$&$    0.0003$\\
G-K93'    &$  -40.8519$&$    0.0653$&$  -38.7983$&$    0.9333$&$    0.0533$\\
G-S55     &$  -41.1342$&$    0.0724$&$  -38.8548$&$    0.9267$&$    0.0592$\\
\hline
\end{tabular}
\label{Tab:cha}
\end{table}

\begin{table}
\small
\centering
\caption{Conversion coefficients between $SFR$ and $L_{\rm 1500}$ (i.e., $i=1$ in Eqs.~\ref{Eq.uv-c1} and ~\ref{Eq.uv-cx}) for all models except for Models Cr-Cha03, Cr-S55, E-K01 and E-S55'.}
\begin{tabular}{lcc ccc}
\hline
  Model & \multicolumn{2}{c}{Eq.~\ref{Eq.uv-c1}} & \multicolumn{3}{c}{Eq.~\ref{Eq.uv-cx}} \\
        & (C$_{\rm 1500}$ & $\sigma_{{\rm 1500}}$) & (C$_{\rm 1500}'$ & A$_{\rm 1500}'$ & $\sigma_{{\rm 1500}}'$)\\
\hline
A-MS79    &$  -28.0681$&$    0.0219$&$  -28.4138$&$    1.0193$&$    0.0176$\\
B-MS79    &$  -28.0037$&$    0.0212$&$  -28.3104$&$    1.0171$&$    0.0178$\\
A-S55     &$  -27.8032$&$    0.0302$&$  -28.2842$&$    1.0272$&$    0.0242$\\
B-S55     &$  -27.7751$&$    0.0305$&$  -28.2403$&$    1.0263$&$    0.0249$\\
C-Cha03   &$  -28.1430$&$    0.0207$&$  -28.4825$&$    1.0193$&$    0.0154$\\
C-S55     &$  -27.9437$&$    0.0232$&$  -28.3225$&$    1.0218$&$    0.0172$\\
D-MS79    &$  -28.1410$&$    0.0208$&$  -28.4820$&$    1.0194$&$    0.0154$\\
F-K93'    &$  -27.7758$&$    0.0092$&$  -27.8092$&$    1.0019$&$    0.0091$\\
F-S55     &$  -27.8955$&$    0.0059$&$  -27.9120$&$    1.0009$&$    0.0059$\\
G-K93'    &$  -27.9137$&$    0.0635$&$  -27.7433$&$    0.9903$&$    0.0630$\\
G-S55     &$  -27.9968$&$    0.0413$&$  -27.6359$&$    0.9796$&$    0.0380$\\
\hline
\end{tabular}
\label{Tab:cl1500}
\end{table}

\begin{table}
\small
\centering
\caption{Conversion coefficients between $SFR$ and $L_{\rm 2800}$ (i.e., $i=2$ in Eqs.~\ref{Eq.uv-c1} and ~\ref{Eq.uv-cx}) for all models except for Models Cr-Cha03, Cr-S55, E-K01 and E-S55'.}
\begin{tabular}{lrr lrr}
\hline
  Model & \multicolumn{2}{c}{Eq.~\ref{Eq.uv-c1}} & \multicolumn{3}{c}{Eq.~\ref{Eq.uv-cx}} \\
        & (C$_{\rm 2800}$ & $\sigma_{{\rm 2800}}$) & (C$_{\rm 2800}'$ & A$_{\rm 2800}'$ & $\sigma_{{\rm 2800}}'$)\\
\hline
A-MS79    &$  -28.0227$&$    0.0674$&$  -29.2780$&$    1.0701$&$    0.0504$\\
B-MS79    &$  -27.9737$&$    0.0714$&$  -29.2873$&$    1.0735$&$    0.0540$\\
A-S55     &$  -27.7947$&$    0.0984$&$  -29.6921$&$    1.1073$&$    0.0729$\\
B-S55     &$  -27.7721$&$    0.1031$&$  -29.7416$&$    1.1115$&$    0.0773$\\
C-Cha03   &$  -28.0637$&$    0.0538$&$  -28.9987$&$    1.0534$&$    0.0390$\\
C-S55     &$  -27.8717$&$    0.0594$&$  -28.9053$&$    1.0597$&$    0.0428$\\
D-MS79    &$  -28.0625$&$    0.0539$&$  -28.9996$&$    1.0535$&$    0.0391$\\
F-K93'    &$  -27.7540$&$    0.0759$&$  -29.1760$&$    1.0805$&$    0.0569$\\
F-S55     &$  -27.8251$&$    0.0459$&$  -28.6379$&$    1.0458$&$    0.0351$\\
G-K93'    &$  -27.9478$&$    0.1704$&$  -30.5344$&$    1.1465$&$    0.1396$\\
G-S55     &$  -27.9483$&$    0.0963$&$  -29.1312$&$    1.0670$&$    0.0832$\\
\hline
\end{tabular}
\label{Tab:cl2800}
\end{table}

\label{Sect:rst-bi}
For the sake of clarity, we define eight sets of models. The two sets of Models A and B use the {\hf Yunnan} EPS models with and without binary interactions, respectively. The six sets of Models C(Cr), D, E, F and G use the {\hf BC03, GISSEL98, PopSTAR, STARBURST99} and {\hf P\'{E}GASE} models, respectively.
In Table~\ref{Tab:defmod} we give the name of each set of models in the 1st column, the name of used EPS model in the 2nd column, the condition that binary interactions are taken into account in the 3rd column, the IMF in the 4th column, and the 5th column denotes the condition that the assumption of gas-recycle is used, where 'N' means $\varepsilon $ in Eq.~\ref{eq.sfr} equals to zero (i.e., the gas could not be recycled into new star formation) and 'Y' means $\varepsilon =1$ (i.e., the gas could be recycled). In the seven sets of Models A-G $\varepsilon = 0$, and in the set of Model Cr $\varepsilon=1$.

For each set of models two subsets are considered, depending on the IMF. To distinguish them, the name of used IMF (the 4th column) is the supplement to the name in the 1st column of Table~\ref{Tab:defmod}.

In this section we mainly discuss the effect of binary interactions on the SFR calibrations. Therefore, only Models A-MS79 and B-MS79 are used, which are based on the standard {\hf Yunnan} EPS models. The other models will be used in the next section to discussion the effects of IMF, gas-recycle assumption and EPS models on the results.

In this section we obtain the number of ionizing photons $Q$(H), the luminosity of H$\alpha$ recombination line $L_{\rm H\alpha}$, the luminosity of [OII]$\lambda$3727 forbidden-line doublet $L_{\rm [OII]}$, the UV fluxes at 1500 and 2800\,$\rm \AA$, $L_{i,{\rm UV}}$, and FIR flux $L_{\rm FIR}$ of Burst, E, S0, Sa-Sd and Irr galaxies for both Models A-MS79 and B-MS79. We also give the conversion factors between SFR and these SFR diagnostics.

\subsection{SFR.vs.$L_{\rm H\alpha}$}
\label{Sect:sfr-Lha}
In Fig.~\ref{Fig:sfr-lha} we give the relation between log($SFR$) and log($L_{\rm H\alpha}$) (note the logarithmic scale) of E, S0, Sa-Sc and Sd galaxy types in the range of 0.1\,Myr-15\,Gyr for Models A-MS79 and B-MS79. Also shown are the SFR($L_{\rm H\alpha}$) calibrations of K98 and B04. If the aperture effect is neglected, the conversion factor of \citet{mat07} is similar to that of B04 (see Section~\ref{Sect:rst-cp}), so we do not show it in Fig.~\ref{Fig:sfr-lha} for the sake of clarity.

From Fig.~\ref{Fig:sfr-lha} we see that the log($SFR$) varies linearly with log($L_{\rm H\alpha}$) within a certain age range (from log$t$/yr$>$7.5 for E-type to log$t$/yr$>$8.5 for Sd-type in Model A-MS79, from log$t$/yr$>$7.1 for E-type to log$t$/yr$>$8.5 for Sd-type in Model B-MS79) for both Models A-MS79 and B-MS79.
And, for an given log($L_{\rm H\alpha}$), the log($SFR$) of Model A-MS79 is smaller by an amount of $\sim$0.2\,dex than that of Model B-MS79 when log$(SFR) \ga -11$, i.e., the inclusion of binary interactions can lower the derived $SFR$ in terms of $L_{\rm H\alpha}$.
When comparing with the SFR of K98 and B04 at a given $L_{\rm H\alpha}$, we find that the log($SFR$) increases from B04, K98, Model A-MS79 and Model B-MS79 in turn, and the discrepancy in the log($SFR$) between K98 and Model A-MS79 is small.

The reason that binary interactions lower the $SFR$ at a given $L_{\rm H\alpha}$ is that they raise the number of ionizing photons $Q$(H). In Fig.~\ref{Fig:qh-t} we give the evolution of log($Q$(H)) for Burst, E, S0-Sc and Sd galaxy types. From it we see that the log($Q$(H)) of Model A-MS79 is significantly greater than that of Model B-MS79 for Bursts in the age range 6.7 $\la$ log\,$t$/yr $\la$ 8.4, the difference in log($Q$(H)) reaching $\sim$2\,dex at an age of log$t$/yr=7.5. For E-S0-Sd galaxy types binary interactions also can raise log($Q$(H)) when log$t$/yr $\ga$ 6.7.

The raise of $Q$(H) is caused by the fact that binary interactions increment the UV spectrum at the corresponding ages. In Fig.~\ref{Fig:ised-t36} we give the stellar and nebular spectra of the Burst galaxy type (SP) at an age of log$t$/yr=7.5 for both Models A-MS79 and B-MS79. From it we see that binary interactions not only raise the stellar spectrum in the UV band, but also raise the nebular continuum (by an amount of  $\sim$2\,dex).

The higher UV spectrum is caused by more hotter stars produced by binary interactions. In Fig.~\ref{Fig:cmd} we give the distribution of stars in log$T_{\rm eff}$-log$g$ plane for Bursts (SPs) at an age of log$t$/yr = 7.2 for Models A-MS79 and B-MS79. From the left panel, we can see that binary interactions (i.e., Model A-MS79) indeed produce some hotter helium stars (log$T_{\rm eff} \sim 4.9$ and log $g \sim 5.8$).
These helium stars are originated from those binary systems, for which the primaries have evolved from the giant branch (GB, core-helium-burning) phase to the helium main-sequence (HeMS) phase due to the mass loss of the primaries. In Fig.~\ref{Fig:evo} we give the evolution of such a binary system from zero age main sequence (ZAMS) to the end of helium burning in the log$T_{\rm eff}$-log$L$ plane. From it we see that the primary would evolve from GB to HeMS phase due to the mass loss, and the luminosity of the secondary increases rapidly due to the mass increase.
In general, these binary systems have relatively large mass ratio $q$ ($>$0.2) and small orbital period ($P$).

To quantitatively analyze the effect of binary interactions on the calibration factor between $SFR$ and $L_{\rm H\alpha}$, we give a fitting relation between log($SFR$) and log($L_{\rm H\alpha}$) when log($SFR) \ge -11$ and $\Delta {\rm log}(SFR) \ge 0.05$ for Models A-MS79 and B-MS79 by the following form:
\begin{equation}
{\rm log} {SFR_{\rm H\alpha} \over (\rm M_\odot \ {\rm yr^{-1}})} =
{\rm log} {L_{\rm H\alpha} \over (\rm ergs \ s^{-1})} + {\rm
C_{H\alpha}},
\label{Eq.ha-c1}
\end{equation}
where $SFR_{\rm H\alpha}$ means that it is calculated from $L_{\rm H\alpha}$. The fitting coefficient (${\rm C_{H\alpha}}$) and the rms ($\sigma_{\rm H{\alpha}}$) are given in the 2nd and 3rd columns of Table~\ref{Tab:cha}, respectively. This form of fitting (Eq.~\ref{Eq.ha-c1}) is the same as that of K98, MPD98 and \citet{mat07}. In fact, it is a linear relation between $SFR$ and $L_{\rm H\alpha}$.

From Fig.~\ref{Fig:sfr-lha} we see that $SFR$ does not vary linearly with $L_{\rm H\alpha}$ (it is easily seen when comparing the results of K98 and B04). Therefore, we also adopt the following form:
\begin{equation}
{\rm log} {SFR_{\rm H\alpha} \over (\rm M_\odot \ {\rm yr^{-1}})} = {\rm
A_{H\alpha}'} \times {\rm log} {L_{\rm H\alpha} \over (\rm ergs \
s^{-1})} + {\rm C_{H\alpha}'}.
\label{Eq.ha-cx}
\end{equation}
The fitted coefficients (${\rm A_{H\alpha}'}$ \& ${\rm C_{H\alpha}'}$) and the rms ($\sigma_{\rm H{\alpha}}'$) are given in the 4th-6th columns of Table~\ref{Tab:cha}. In the work of H03, the relation between $SFR$ and the U-band luminosity is fitted by the above form (see their Eq.\,10).
In the top panels of Fig.~\ref{Fig:ha-fit} we give the comparisons in the log($SFR$).vs.log($L_{\rm H\alpha}$) relation between fitting (log $SFR_{\rm fit}$) and model data for Model A-MS79 in both cases of Eqs.~\ref{Eq.ha-c1} and \ref{Eq.ha-cx}.
In the bottom panels of Fig.~\ref{Fig:ha-fit} we give the residuals (log$SFR$-log$SFR_{\rm fit}$) as a function of log$L_{\rm H\alpha}$.

\subsection{SFR.vs.$L_{1500}$ and SFR.vs.$L_{2800}$}
\label{Sect:sfr-Luv}
In Fig.~\ref{Fig:sfr-uv} we give the relations between log($SFR$) and the logarithmic UV-luminosities at 1500 and 2800\,$\rm \AA$ of E, S0-Sd types of galaxies for Models A-MS79 and B-MS79, also give the results of K98, MPD98 and G10. The results of MPD98 are obtained by using S55 with $\alpha=-2.35$ and \citet{sca86} IMF.

From the left and right panels of Fig.~\ref{Fig:sfr-uv}, we see that the differences in the $SFR$-$L_{1500}$ and $SFR$-$L_{2800}$ relations between Models A-MS79 and B-MS79 are small.
By comparing the $SFR$ of K98, MPD98 and G10 at given $L_{1500}$ or $L_{2800}$, we find that the log($SFR$) increase from G10, Models A-MS79/B-MS79, MPD98-S55, K98 and MPD98-Scalo in turn (i.e., $SFR_{\rm G10} > SFR_{\rm A-MS79/B-MS79} >SFR_{\rm MPD98-S55} >SFR_{\rm K98} >SFR_{\rm MPD98-Scalo}$), the differences between MPD98-S55 and K98 are small, and our results lie between G10 and MPD98-S55.

The reason that binary interactions do not affect $SFR$ at given $L_{1500}$ or $L_{2800}$ can be seen from Fig.~\ref{Fig:uv-t}, in which we give the evolutions of $L_{1500}$ and $L_{2800}$ of Burst, E, S0-Sc and Sd types of galaxies for Models A-MS79 and B-MS79.
First, from the left panel of Fig.~\ref{Fig:uv-t}, we see that the $L_{1500}$ of Model A-MS79 is greater than that of Model B-MS79 only for Bursts at ages of log$t$/yr $\sim$ 9 (the maximal difference is $\sim$1\,dex at an age of 1\,Gyr). For the other galaxy types the difference in the $L_{1500}$ is small. The
reason for this is that the contribution of the SP with an age of $\sim$1\,Gyr to the ISEDs is smaller for E, S0-Sd galaxies according to
\begin{equation}
F(t) = \int_{0}^{t} \psi(t-t') f_{\rm SP}(t') {\rm d}t',
\end{equation}
where $F(t)$ is the galaxy spectrum at time $t$, $\psi(t-t')$ is the SFR at $t-t'$ (see Eq.~\ref{eq.sfr}) and $f_{\rm SP}(t')$ is the flux of SP with an age of $t'$. In addition, from the right panel we see that the difference in the $L_{2800}$ between Models A-MS79 and B-MS79 is insignificant for all galaxy types at all ages.
Therefore, binary interactions would not affect significantly the $SFR$-$L_{1500}$ and $SFR$-$L_{2800}$ relations.

Also we fit the log($SFR$)-log($L_{1500}$) and log($SFR$)-log($L_{2800}$) relations by using the expressions
\begin{equation}
{\rm log} {SFR_ {i, {\rm UV}} \over (\rm M_\odot \ {\rm yr^{-1}})} =
{\rm log} {L_{i, {\rm UV}} \over (\rm ergs \ s^{-1} \ Hz^{-1})} + {\rm
C}_{i, {\rm UV}}
\label{Eq.uv-c1}
\end{equation}
and
\begin{equation}
{\rm log} {SFR_ {i, {\rm UV}} \over (\rm M_\odot \ {\rm yr^{-1}})} =
{\rm A}_{i, {\rm UV}}' \times {\rm log} {L_{i, {\rm UV}} \over (\rm ergs
\ s^{-1} \ Hz^{-1})} + {\rm C}_{i, {\rm UV}}',
\label{Eq.uv-cx}
\end{equation}
where $i$=1 denotes the wavelength $\lambda= 1500\,\rm \AA$, $i$=2 means $\lambda = 2800\,\rm \AA$, and $SFR_{i, {\rm UV}}$ means that it is from the $i$-th UV luminosity $L_{i, {\rm UV}}$. The fitting coefficients between log($SFR$) and log($L_{1500}$) (${\rm C}_{i, {\rm UV}}, {\rm C}_{i, {\rm UV}}', {\rm A}_{i, {\rm UV}}'$) and rms ($\sigma_{i, {\rm UV}}, \sigma_{i, {\rm UV}}'$) are given in Table~\ref{Tab:cl1500}, and those between log($SFR$) and log($L_{2800}$) are given in Table~\ref{Tab:cl2800}.

\subsection{SFR.vs.$L_{\rm [OII]}$}
\label{Sect:sfr-Loii}
The luminosity of the [OII]$\lambda3727\rm \AA$ forbidden-line doublet, $L_{\rm [OII]}$, is indirectly related to the ionizing luminosity. Often it is obtained by using the ratio of $L_{\rm [OII]}/L_{\rm H\alpha}$, which is obtained empirically. In the work of K98 $L_{\rm [OII]}/L_{\rm H\alpha} = 0.45$, in the work of H03 it is 0.23, in the work of G10 the value is 0.5, and in the {\hf P\'{E}GASE} code the value is 3.01/2.915. In this work we use the value of 0.23 used by H03 to obtain $L_{\rm [OII]}$.
Because the fixed $L_{\rm [OII]}/L_{\rm H\alpha}$ ratio is used, the plot of the $SFR$.vs.$L_{\rm [OII]}$ is similar to Fig.~\ref{Fig:sfr-lha}. At a given $SFR$, the log($L_{\rm [OII]}$) is smaller than log($L_{\rm H\alpha}$) by an amount of log(1/0.23), i.e., the calibration curve moves upwards by log(1/0.23).
Therefore we do not give the plot of $SFR$-$L_{\rm [OII]}$ relation. Combining the conclusion made in Section~\ref{Sect:sfr-Lha} we know that binary interactions make the log($SFR$) in terms of $L_{\rm [OII]}$ larger by about 0.2\,dex at a given $L_{\rm [OII]}$.

Also, we give a fitting relation between $SFR$ and $L_{\rm [OII]}$ by the following two forms:
\begin{equation}
{\rm log} {SFR_{\rm [OII]} \over (\rm M_\odot \ {\rm yr^{-1}})} =  {\rm
log} {L_{\rm [OII]} \over (\rm ergs \ s^{-1})} + {\rm C_{[OII]}}
\label{Eq.oii-c1}
\end{equation}
and
\begin{equation}
{\rm log} {SFR_{\rm [OII]} \over (\rm M_\odot \ {\rm yr^{-1}})} =  {\rm
A_{[OII]}'} \times {\rm log} {L_{\rm [OII]} \over (\rm ergs \ s^{-1})} +
{\rm C_{[OII]}'}.
\label{Eq.oii-cx}
\end{equation}
Because the fixed $L_{\rm [OII]}/L_{\rm H\alpha}$ ratio is used, the fitting coefficient ${\rm A_{[OII]}'} = {\rm A_{L_{H\alpha}}'}$, ${\rm C_{[OII]} = \rm C_{H\alpha}-log(0.23)}$ and ${\rm C'_{[OII]} = \rm C_{H\alpha}'-log(0.23)}$ (cf. Eqs.~\ref{Eq.ha-c1}-\ref{Eq.ha-cx}). The actual numbers can immediately be computed from the values in Table 3.

\subsection{SFR.vs.$L_{\rm FIR}$}
\label{Sect:sfr-Lfir}
In the computations of the above three SFR diagnostics (Section~\ref{Sect:sfr-Lha}-\ref{Sect:sfr-Loii}), the calibration factors are from those models with an exponentially decreasing SFR (i.e., Eq.~\ref{eq.sfr}),
while the calibration between SFR and FIR luminosity is from the models with constant SFR under the assumption of the bolometric luminosity $L_{\rm BOL} = L_{\rm FIR}$.

In Fig.~\ref{Fig:fir-t} we give the $L_{\rm FIR}$ evolution of Irr galaxies (i.e., models with constant SFR) for Models A-MS79 and B-MS79. Also shown are the results of K98. In all cases the total mass of the model galaxy is normalized to 1\,M$_{\rm \odot}$.
From this figure we see that the difference in the $L_{\rm FIR}$ evolution between Models A-MS79 and B-MS79 is small, that is to say, binary interactions almost do not vary the conversion factor between $SFR$ and $L_{\rm FIR}$.

\section{Other factors on SFR calibrations}
\begin{figure}
\centering
\includegraphics[height=8.0cm,width=6.5cm,angle=270]{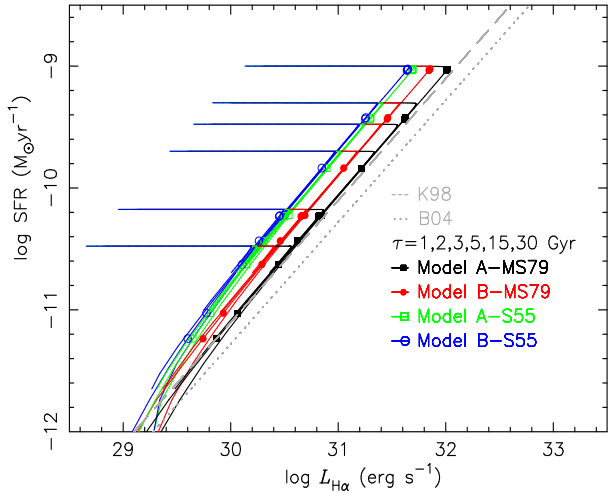}
\caption{Similar to Fig.~\ref{Fig:sfr-lha}, but including the results of Models A-S55 (green solid line+open rectangles) and B-S55 (blue solid line+open circles).}
\label{Fig:sfr-lha-imf}
\end{figure}

\begin{figure}
\includegraphics[height=8.0cm,width=6.5cm,angle=270]{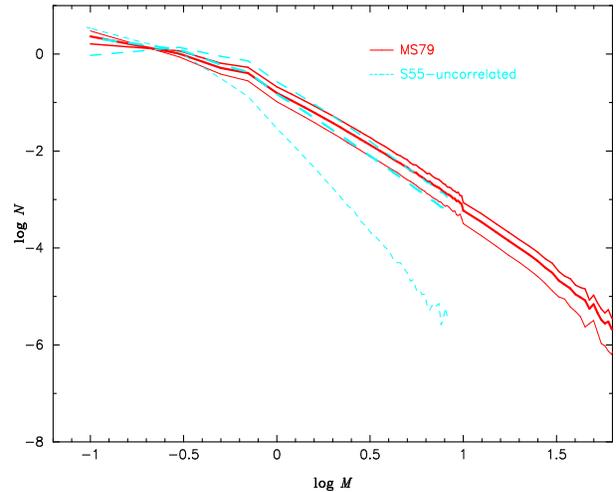}
\centering
\caption{Comparison of the IMF between Models A/B-MS79 (red, solid line) and A/B-S55 (cyan, dashed line). The thick lines are for the mass of binary system, the upper and lower lines are for the masses of the primary and secondary, respectively. Note in this figure log$N$ is obtained by using the EFT's approximation to the MS79 and S55 IMFs.}
\label{Fig:imf-ms79s55}
\end{figure}

\begin{figure}
\centering
\includegraphics[height=8.0cm,width=6.5cm,angle=270]{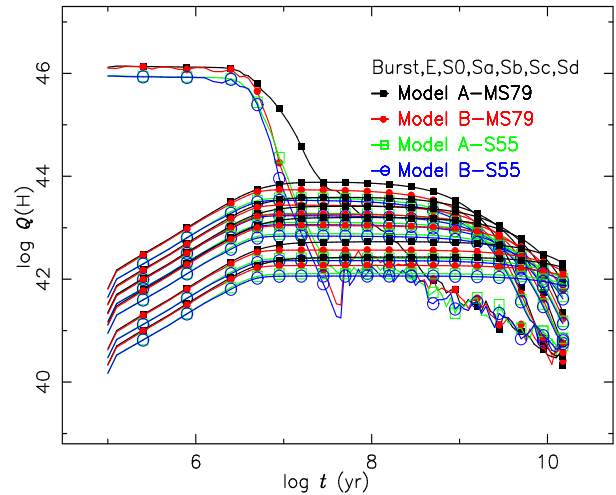}
\caption{Similar to Fig.~\ref{Fig:qh-t}, but also shown are the results of Models A-S55 (green line+open rectangles) and B-S55 (blue line+open circles).}
\label{Fig:qh-t-imf}
\end{figure}

\begin{figure}
\centering
\includegraphics[height=8.0cm,width=6.5cm,angle=270]{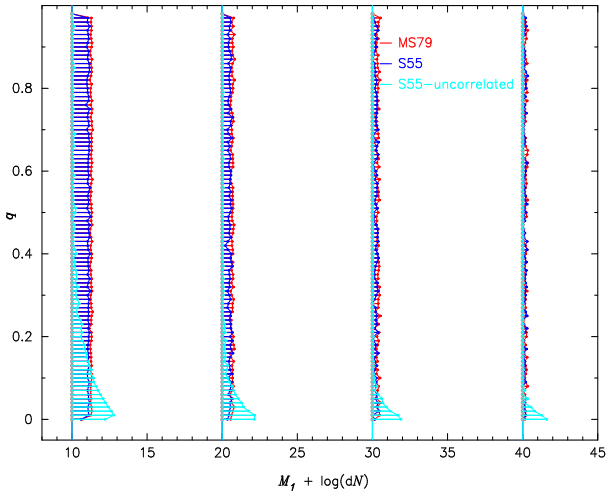}
\caption{The number of binary systems in the ranges $M_1 \rightarrow M_1+ {\rm d}M_1$ and $q \rightarrow q+{\rm d}q$ (d$q$=0.02) per million binary systems (d$N$) for Models A/B-MS79 (red) and A/B-S55 (cyan). The grey circles denote the grid ($M_1, q$), and the distance from grid to the right point denotes the number of binary systems. For comparison, we also give the results of using the S55 IMF and the assumption that the masses of the two component stars are correlated (blue).
}
\label{Fig:imf-ms79s55-q}
\end{figure}

\begin{figure*}
\includegraphics[angle=270,scale=.400]{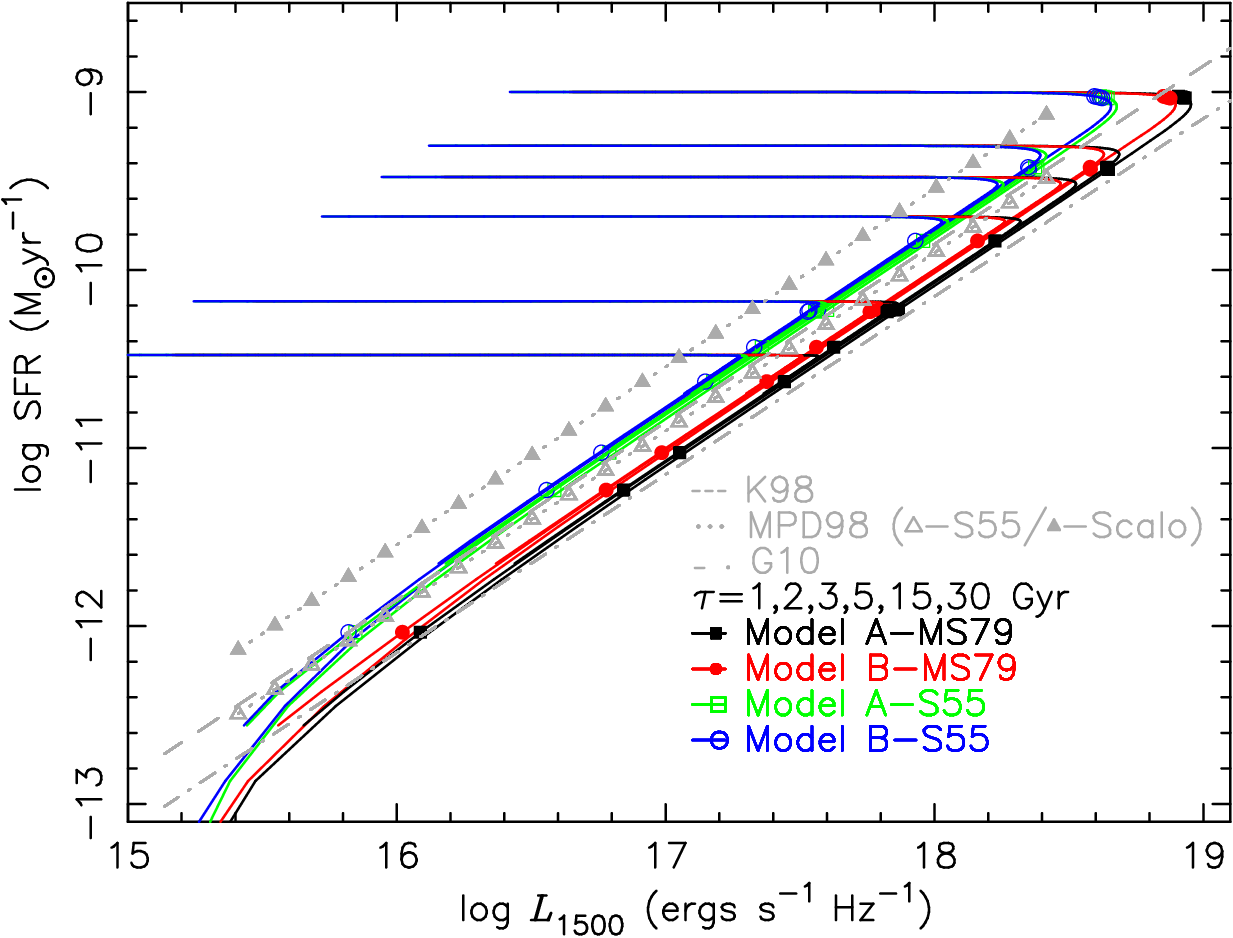}
\includegraphics[angle=270,scale=.400]{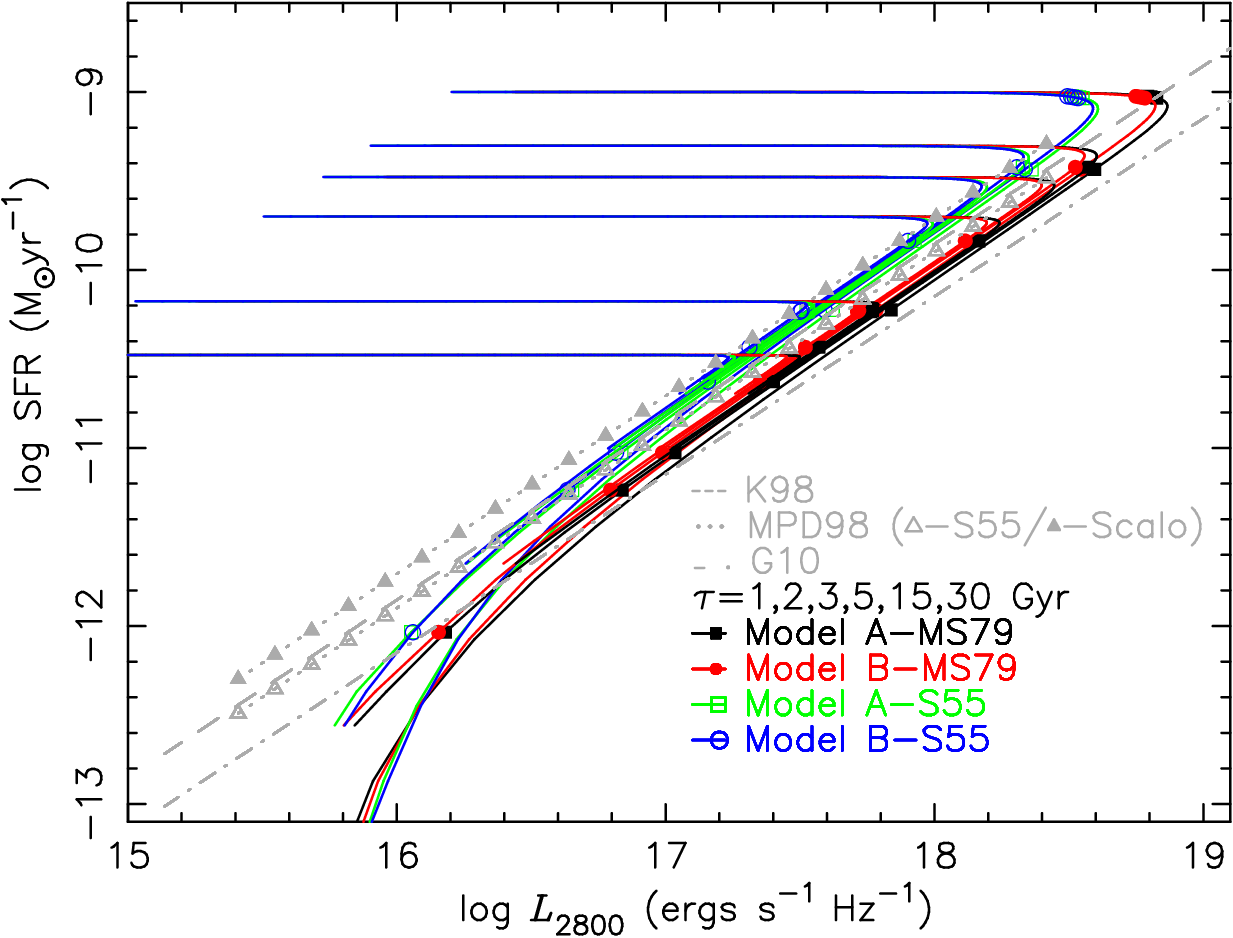}
\caption{Similar to Fig.~\ref{Fig:sfr-uv}, but including the results of
Models A-S55 (green solid line+open rectangles) and B-S55 (blue solid
line+open circles).} \label{Fig:sfr-uv-imf}
\end{figure*}

\begin{figure*}
\includegraphics[angle=270,scale=.400]{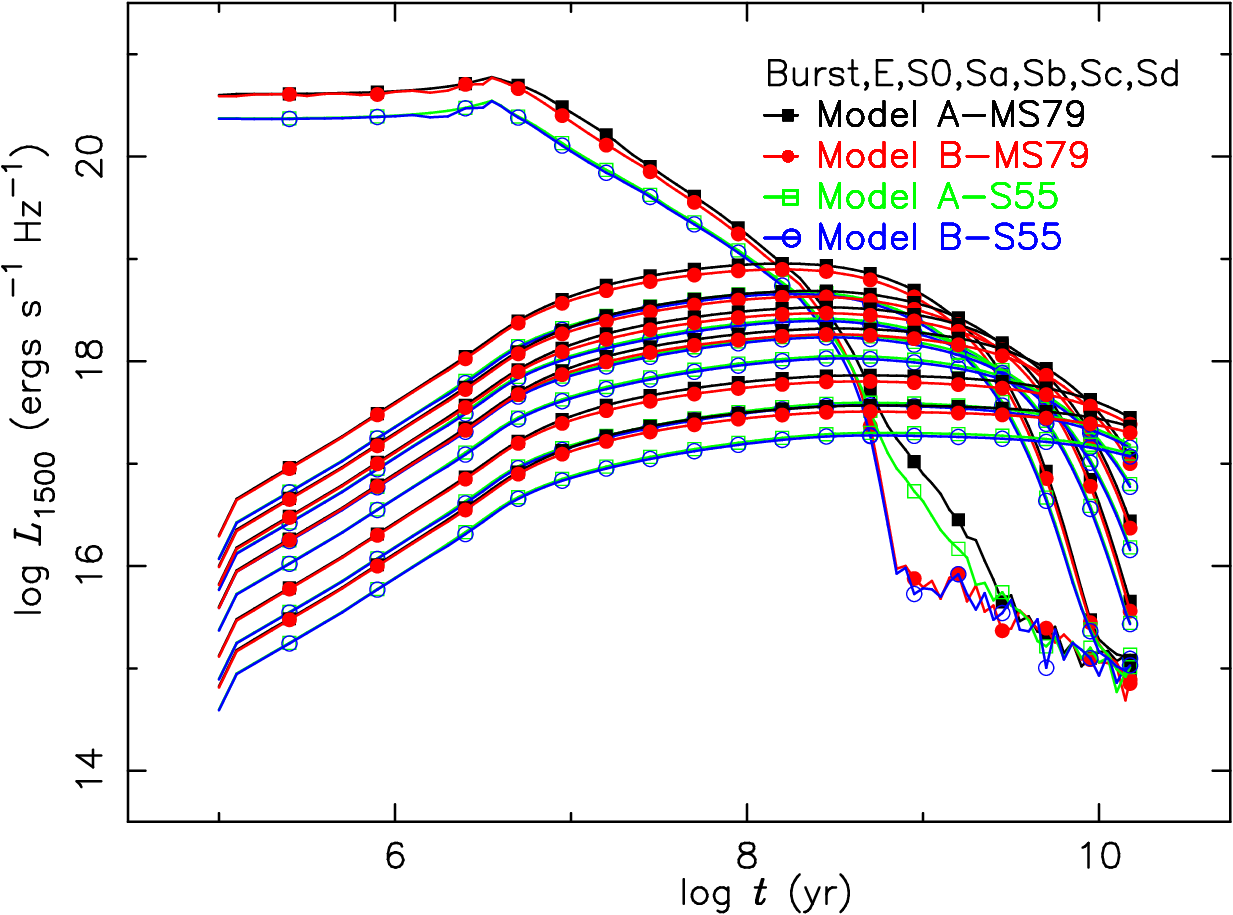}
\includegraphics[angle=270,scale=.400]{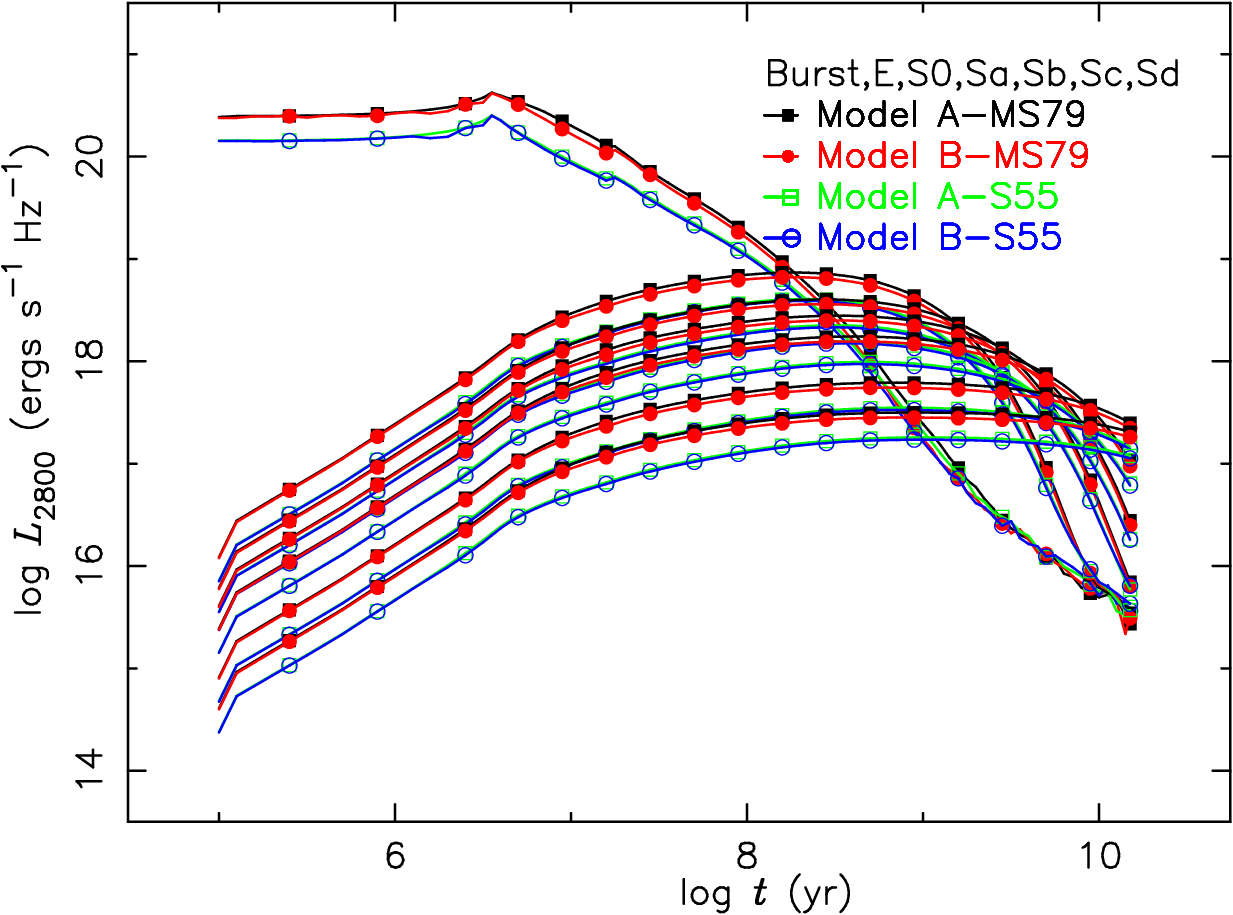}
\caption{Similar to Fig.~\ref{Fig:uv-t}, but including the results of Models A-S55 (green line+open rectangles) and B-S55 (blue line+open circles).}
\label{Fig:uv-t-imf}
\end{figure*}

%
\label{Sect:rst-ft}
The calibrations between $SFR$ and $L_{\rm H\alpha}$, $L_{\rm 1500}$, $L_{\rm 2800}$, $L_{\rm [OII]}$ and $L_{\rm FIR}$, which have been given in Section~\ref{Sect:rst-bi}, can be affected by the adoption of different EPS models, IMF and the assumption of gas recycle.
In this section, we will discuss the effects of these factors on the above mentioned $SFR$ calibrations by introducing the Models A-S55, B-S55, ..., G-K93' and G-S55.
For these models, the $L_{\rm H\alpha}$, $L_{\rm 1500}$, $L_{\rm 2800}$, $L_{\rm [OII]}$ and $L_{\rm FIR}$ are calculated, and the calibration factors are presented in Tables.~\ref {Tab:cha}, ~\ref{Tab:cl1500} and ~\ref{Tab:cl2800}, respectively.

\subsection{The effect of IMF}
\label{Sect:ft-imf}
To analyse the effect of IMF on these calibrations, the results of Models A-S55 and B-S55 are needed to be combined with those of Models A-MS79 and B-MS79 (Section~\ref{Sect:rst-bi}). Models A-S55 and B-S55 are built by using the {\hf Yunnan} models with the S55 IMF and the assumption that the masses of the two component stars in a binary system are uncorrelated.

\subsubsection{IMF on $SFR$.vs.$L_{\rm H\alpha}$}
\label{Sect:ft-imf-Lha}
In Fig.~\ref{Fig:sfr-lha-imf} we give the comparison in the $SFR$-$L_{\rm H\alpha}$ relation among Models A-MS79, A-S55, B-MS79 and B-S55. Also shown are the results of K98 and B04.

By comparing the $SFR$-$L_{\rm H\alpha}$ relation between Models A-MS79 and A-S55, and that between Models B-MS79 and B-S55, we see that the log($L_{\rm H\alpha}$) of Model A-S55 is smaller than that of Model A-MS79 by an amount of 0.4\,dex, and that the prediction for Model B-S55 is smaller by $\sim$0.2\,dex than that of Model B-MS79 for a given log($SFR$).
This is partly caused by the fact that Models A/B-S55 produce less massive-stars (see  Fig.~\ref{Fig:imf-ms79s55}) and less $Q$(H) (see Fig.~\ref{Fig:qh-t-imf}).
In Fig.~\ref{Fig:imf-ms79s55} we present a comparison of the IMF between Models A/B-MS79 and A/B-S55, where the S55 and MS79 IMFs are obtained by using the EFT's approximation (i.e., Eqs.~\ref{eq.imfms79-app} and \ref{eq.imfs55-app}) and are different from those given by Eqs.~\ref{eq.imfms79} and \ref{eq.imfs55}.
In Fig.~\ref{Fig:qh-t-imf} we present the $Q$(H) evolution of all galaxy types for Models A-MS79, B-MS79, A-S55 and B-S55, respectively.
From them we indeed see that the number of massive stars and therefore $Q$(H) at early ages of Bursts for Models A/B-S55 is less than those for Models A/B-MS79.

Moreover, by comparing the result between Models A-S55 and B-S55 we find that the difference in the $SFR$-$L_{\rm H\alpha}$ relation between them is small, which is significantly different from that between Models A-MS79 and B-MS79.
The difference in the $SFR$-$L_{\rm H\alpha}$ relation between Models A-MS79 and B-MS79 is caused by the difference in the $Q$(H) of Bursts in the age range 6.7 $\la$ log$t$/yr $\la$ 8.4,
while the difference in the $Q$(H) between Models A-S55 and B-S55 is small for Bursts at all ages (see Fig.~\ref{Fig:qh-t-imf}). Why Model A-S55 could not produce more $Q$(H) than Model B-S55, like Model A-MS79?

This is because less hotter helium stars (log$T_{\rm eff}\sim 4.9$ and log$ g \sim 5.8$) could be produced in the range 6.7 $\la$ log$t$/yr $\la$ 8.4 for Model A-S55 (see the right panel of Fig.~\ref{Fig:cmd}, and the comparison with the corresponding left-right panel).
We can understand it from the discussion in Section~\ref{Sect:sfr-Lha} and from Figs.~\ref{Fig:imf-ms79s55} and~\ref{Fig:imf-ms79s55-q}.
First, from the discussion in the Section~\ref{Sect:sfr-Lha} we know that those hotter helium stars (present in the left panel of Fig.~\ref{Fig:cmd}) evolve from those initial binary systems with the relatively large primary-mass ($M_1$), large mass-ratio ($q$) and small orbital separation.
Then, from Fig.~\ref{Fig:imf-ms79s55} we know that the number of massive primary-stars becomes to be less for Model A-S55. Furthermore, from Fig.~\ref{Fig:imf-ms79s55-q} we know that the number of binary systems with relatively large $M_1$ and large $q$ becomes to be less for Model A-S55.
In Fig.~\ref{Fig:imf-ms79s55-q} we give the the number of binary systems in the primary-mass range $M_1 \rightarrow M_1 + {\rm d}M_1$ and the mass-ratio range $q \rightarrow q+{\rm d}q$ for Models A/B-MS79 and A/B-S55. The results of using the S55 IMF and assuming that the masses of the two component stars are correlated (the same as that in Models A/B-MS79) are also represented.
For the sake of clarity, in Fig.~\ref{Fig:imf-ms79s55-q} we only give the number of binaries with the primary-mass 8 $\le M_1 \le 45 {\rm M_\odot}$. These binaries are likely to evolve into hotter helium stars in the range 6.7 $\la$ log$t$/yr $\la$ 8.4, and the MS lifetimes of single stars with the upper and lower masses (8 \& 45 ${\rm M_\odot}$) approximately correspond to the ages of log$t$/yr=6.7 and 8.4 (the MS lifetime of $M=3.5\rm{M_\odot}$ is log$t$=8.4, the MS lifetime of star with $M=38.\rm{M_\odot}$ is log$t$=6.7 at solar metallicity).
From it we see that the number of binary systems with larger $M_1$ and $q$ is less for Model A-S55. Thus Model A-S55 could not produce more hotter helium stars and $Q$(H).
Moreover, from Fig.~\ref{Fig:imf-ms79s55-q} we also see that the number of binaries with larger $M_1$ and $q$ of Model A-MS79 is similar to that by using S55 IMF and assuming that the masses of the two component stars are correlated. Therefore,
the real reason why Model A-S55 could not produce hotter helium stars is the assumption about the masses of two component stars in binary systems.

From the above discussion, we conclude that the $SFR$.vs.$L_{\rm H\alpha}$ calibration is affected not only by the adoption of the different IMF, but also by the assumption about the masses of the component stars in binary systems.

\subsubsection{IMF on $SFR$.vs.$L_{\rm 1500}$ and $L_{\rm 2800}$}
\label{Sect:ft-imf-Luv}
In Fig.~\ref{Fig:sfr-uv-imf} we give the comparisons in the $SFR$-$L_{1500}$ and $SFR$-$L_{\rm 2800}$ relations among Models A-MS79, A-S55, B-MS79 and B-S55. Also shown are the results of K98, MPD98 and G10.

By comparisons, we find that the log($L_{1500}$) and log($L_{\rm 2800}$) of Models A/B-S55 are smaller than the corresponding ones of Models A/B-MS79 by about 0.2\,dex at a given log($SFR$), the conversion curves move upwards from the lines of K98 and MPD98-S55.
For the reason we provided in Section~\ref{Sect:ft-imf-Lha} (less massive stars), and from the results displayed in Fig.~\ref{Fig:uv-t-imf} (which represents the UV luminosities for all galaxy types predicted by the Models A-MS79, B-MS79, A-S55 and B-S55).
We see that the $L_{1500}$ and $L_{2800}$ of Models A/B-S55 are smaller than the corresponding ones for Models A/B-MS79 at all ages for Bursts, leading to smaller $L_{1500}$ and $L_{2800}$ for E, S0-Sd galaxies and therefore larger conversion factors between $SFR$ and $L_{1500}$ and between $SFR$ and $L_{2800}$.

From the left panel of Fig.~\ref{Fig:sfr-uv-imf}, we see that the difference in the $SFR$-$L_{1500}$ relation between Models A-S55 and B-S55 is small, which is similar to the case of using MS79 IMF (see Fig.~\ref{Fig:sfr-uv}).
The reason is the same given in the discussion presented in the 3rd paragraph of Section~\ref{Sect:sfr-Luv}, i.e., the $L_{1500}$ of Models A-S55 is smaller than that of B-S55 only for Bursts in the age range 8.75 $\la$ log$t$/yr $\la$ 9.2,  and the maximal difference between them is $\sim$1\,dex (also see the left panel of Fig.~\ref{Fig:uv-t-imf}). This would not affect significantly the $L_{1500}$ of E, S0-Sd galaxies. Therefore IMF would not affect significantly the $SFR$.vs.$L_{1500}$ conversion.

Moreover, from the right panel of Fig.~\ref{Fig:sfr-uv-imf}, we see that the difference in the $SFR$-$L_{2800}$ relation between Models A-S55 and B-S55 is also small. This is because that IMF does not affect the $L_{2800}$ for Bursts at all ages.

\subsubsection{IMF on $SFR$.vs.$L_{\rm [OII]}$}
\label{Sect:ft-imf-Loii}
Because we use the fixed $L_{\rm [OII]}/L_{\rm H\alpha}$ ratio, the effect of IMF on the $SFR$.vs.$L_{\rm [OII]}$ conversion is the same as that on the $SFR$.vs.$L_{\rm H\alpha}$ conversion, i.e., the conversion factor between $SFR$ and $L_{\rm [OII]}$ of Model A-S55 is larger than that of Model A-MS79 by $\sim$0.4\,dex, the conversion factor of Model B-S55 is larger than that of Model B-MS79 by $\sim$0.2\,dex, and the difference in the $SFR$.vs.$L_{\rm [OII]}$ conversion between Models A-S55 and B-S55 is small.

\subsubsection{IMF on $SFR$.vs.$L_{\rm FIR}$}
In Fig.~\ref{Fig:fir-t} we give the $L_{\rm FIR}$ evolution of Irr galaxies for Models A-S55 and B-S55.
From it we see that the $L_{\rm FIR}$ of Models A/B-S55 is lower than that of Models A/B-MS79 by an amount of $\sim$0.2\,dex. Therefore, the conversion factor between $SFR$ and $L_{\rm FIR}$ of Models A/B-S55 is larger than that of Models A/B-MS79 by an amount of $\sim$0.2\,dex.

\subsection{The effect of gas-recycle assumption}
\begin{figure*}
\centering
\includegraphics[angle=270,scale=.400]{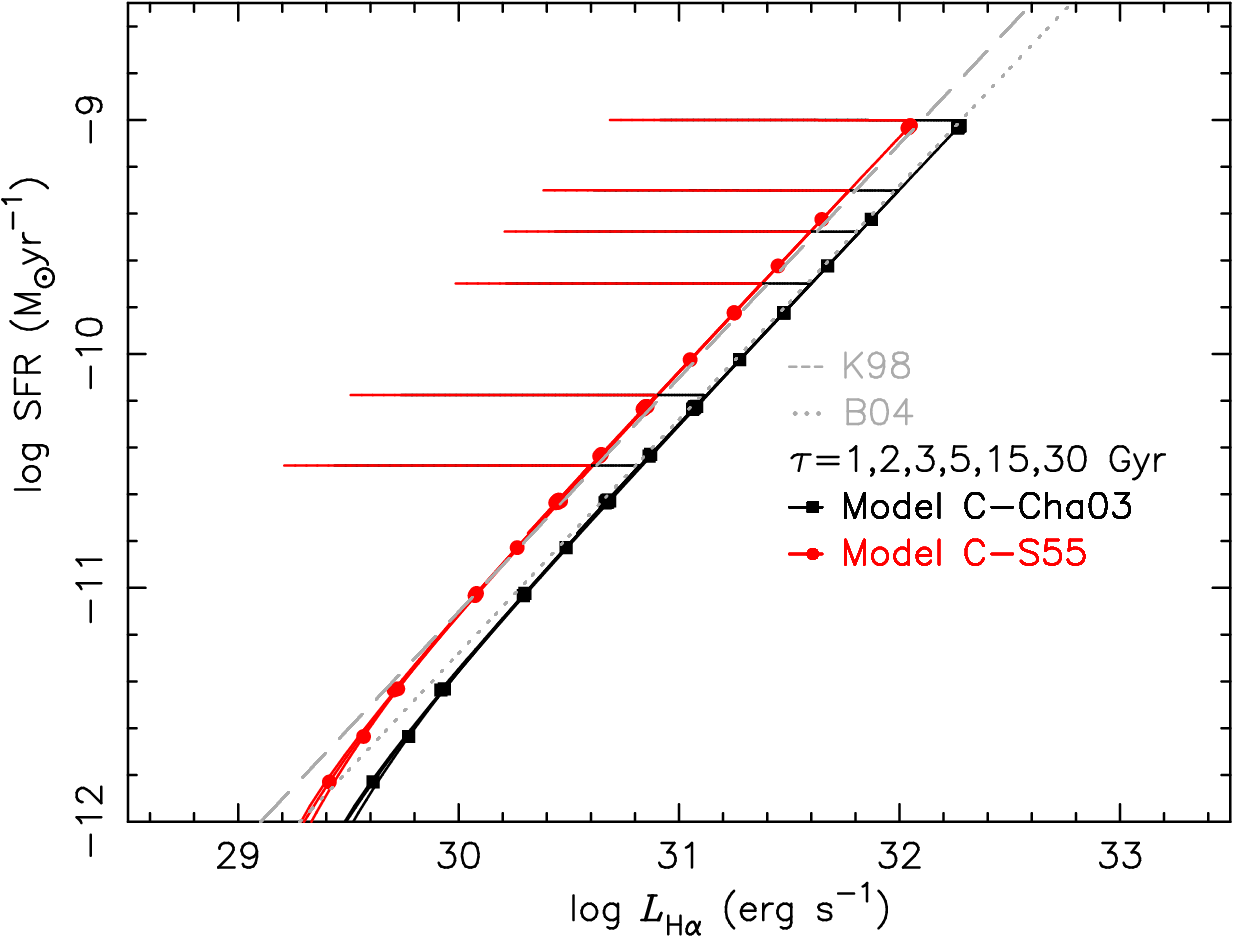}
\includegraphics[angle=270,scale=.400]{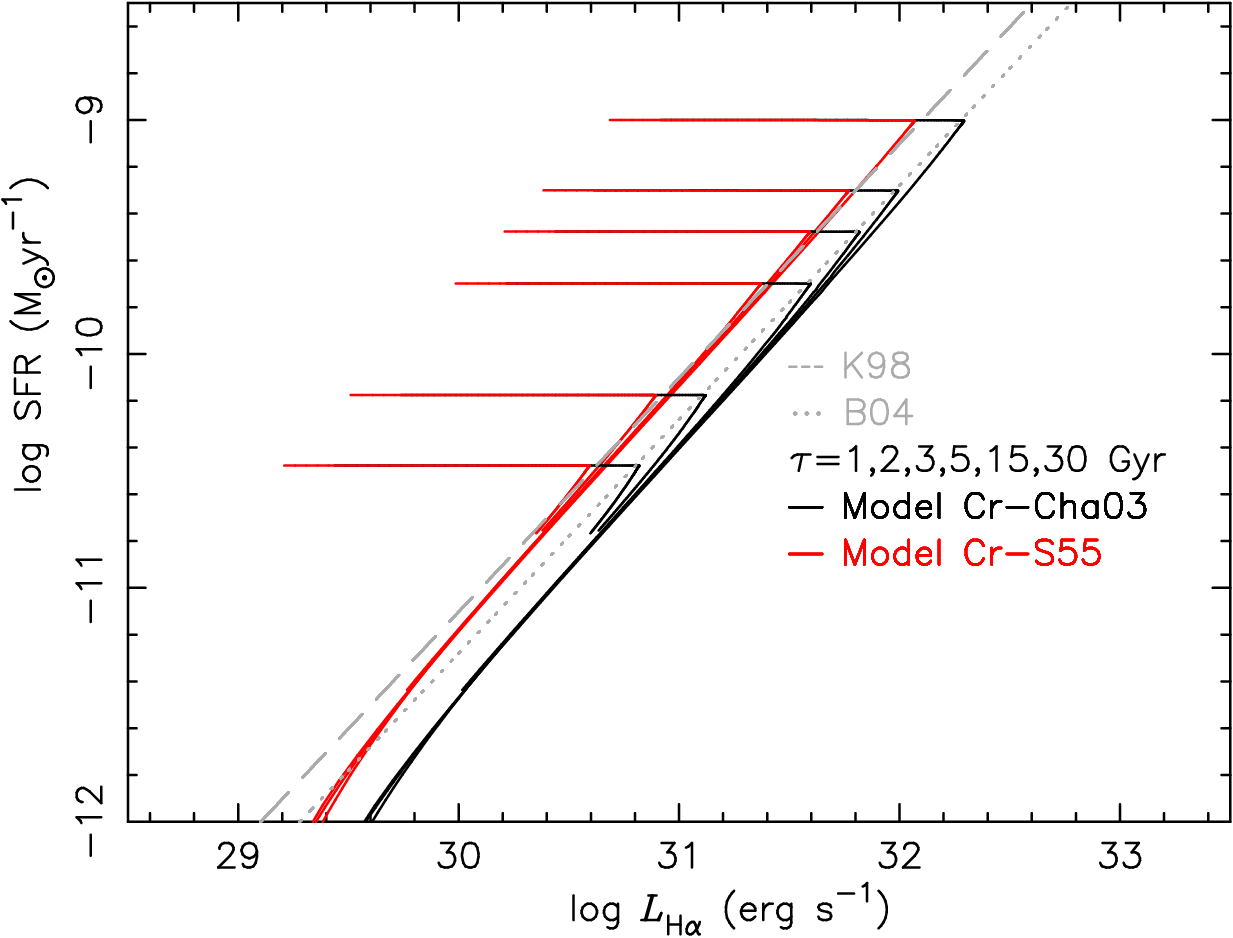}
\caption{Similar to Fig.~\ref{Fig:sfr-lha}, but for Models C (Cha03 and S55 IMFs, left panel) and Cr (Cha03 and S55 IMFs, right panel).}
\label{Fig:sfr-lha-gas}
\end{figure*}

\begin{figure}
\centering
\includegraphics[height=8.0cm,width=6.5cm,angle=270]{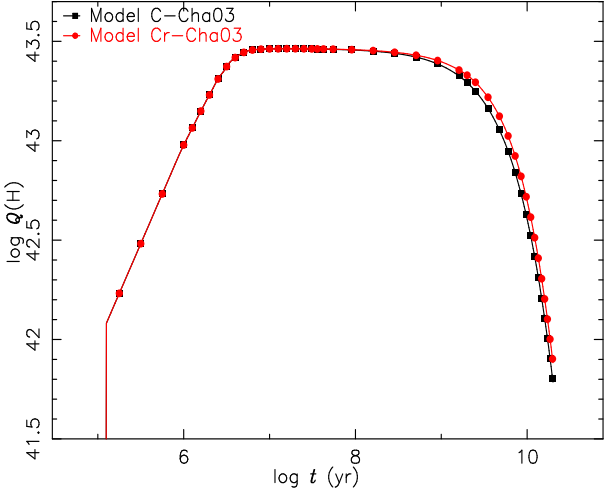}
\caption{Similar to Fig.~\ref{Fig:qh-t}, but for Models C-Cha03 (black line+solid rectangles) and Cr-Ch03 (red line+solid circles) and only for Sb-type galaxies.}
\label{Fig:qh-t-gr}
\end{figure}

In this part we use the Models C and Cr to discuss the effect of gas-recycle assumption on these $SFR$ calibrations. In Models C and Cr the value of $\epsilon$ (see Eq.~\ref{eq.sfr}) is set to 0 and 1, respectively, i.e., in Model C the gas can not be recycled into new star formation, while in Model Cr it can.

In Fig.~\ref{Fig:sfr-lha-gas} we give the calibration between $SFR$ and $L_{\rm H\alpha}$ for Models C-Cha03, C-S55, Cr-Cha03 and Cr-S55 (left and right are for Models C and Cr, respectively).
By comparison, we see that log($SFR$) does not vary linearly with log($L_{\rm H\alpha}$) within a certain SFR (age) range for E, S0-Sd types of galaxies when considering the gas-recycle assumption (i.e, Models Cr-Cha03/S55), and that log($L_{\rm H\alpha}$) of Models Cr-Cha03/S55 is greater than that of Models C-Cha03/S55 at a given log$(SFR)$ (easily seen by comparing with the line of B04).
This is because the inclusion of the gas-recycle assumption produces more $Q$(H) and larger $L_{\rm H\alpha}$.
In Fig.~\ref{Fig:qh-t-gr} we give the evolution of $Q$(H) of Sd-type galaxies for Models C-Cha03 and Cr-Cha03.

Moreover, from Fig.~\ref{Fig:sfr-lha-gas} we see that the difference in the log($SFR$)-log($L_{\rm H\alpha}$) relation between Models C and Cr increases with decreasing SFR (increasing age).
For example, the line of Model C-Cha03 overlaps that of B04 at log$(SFR) \sim-11$, while that of Cr-Cha03 is shifted to the right of the line of B04 ($\sim$0.1\,dex).
The reason for this is that the difference in the $Q$(H) increases with age (see Fig.~\ref{Fig:qh-t-gr}).

For the $L_{\rm 1500}$ and $L_{\rm 2800}$, the situation is similar to that of $L_{\rm H\alpha}$. For the sake of size, we do not give the plots for them in this paper. The inclusion of gas-recycle assumption can lower the derived $SFR$ in terms of $L_{i,{\rm UV}}$ by an amount of 0.05\,dex at log$(SFR) \sim-12.$

For the $L_{\rm [OII]}$, the situation is the same as that of $L_{\rm H\alpha}$. As for the $L_{\rm FIR}$, the inclusion of gas-recycle assumption almost does not affect the conversion factor between $SFR$ and $L_{\rm FIR}$.

\subsection{The effect of EPS models}
\begin{figure*}
\centering
\includegraphics[angle=270,scale=.400]{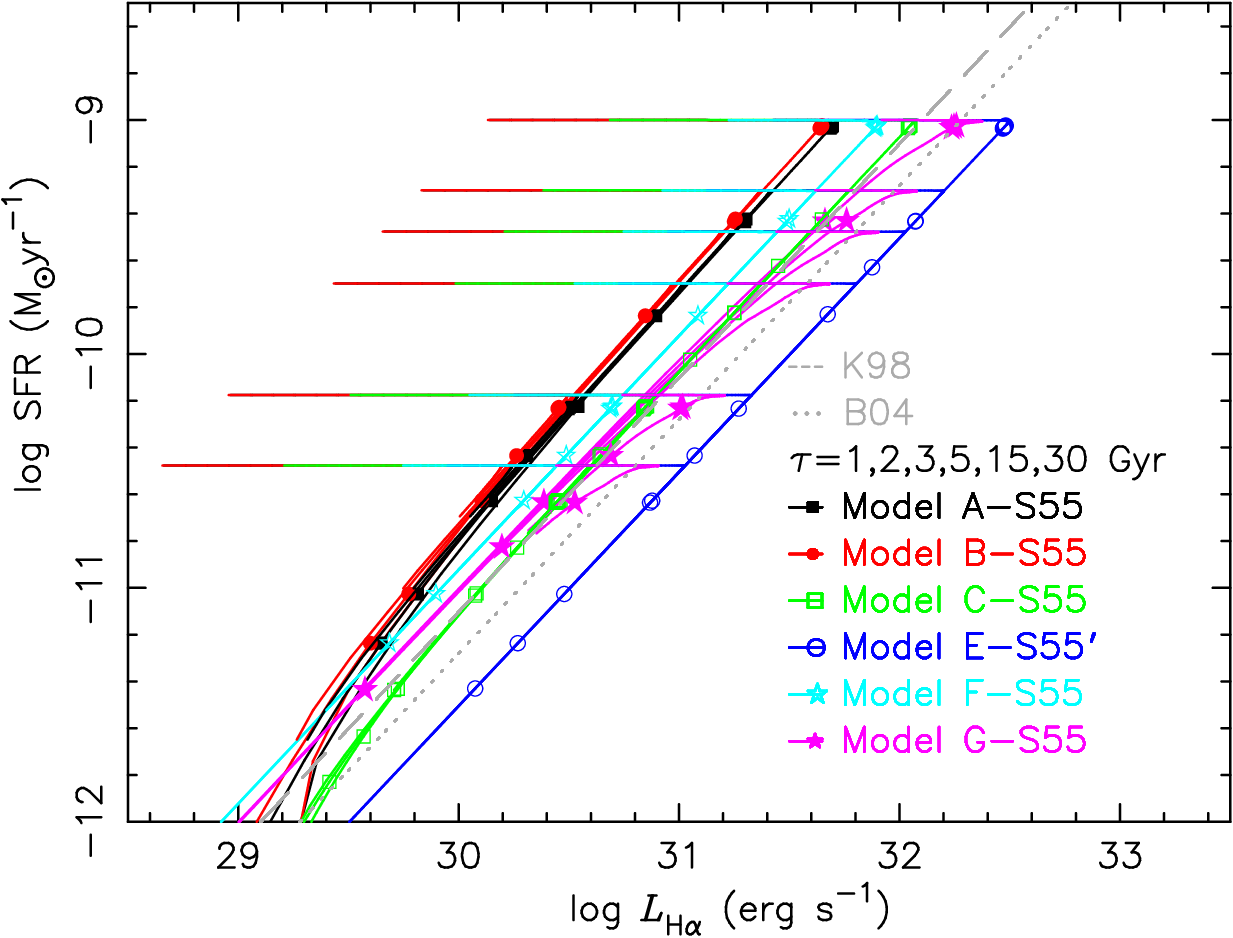}
\includegraphics[angle=270,scale=.400]{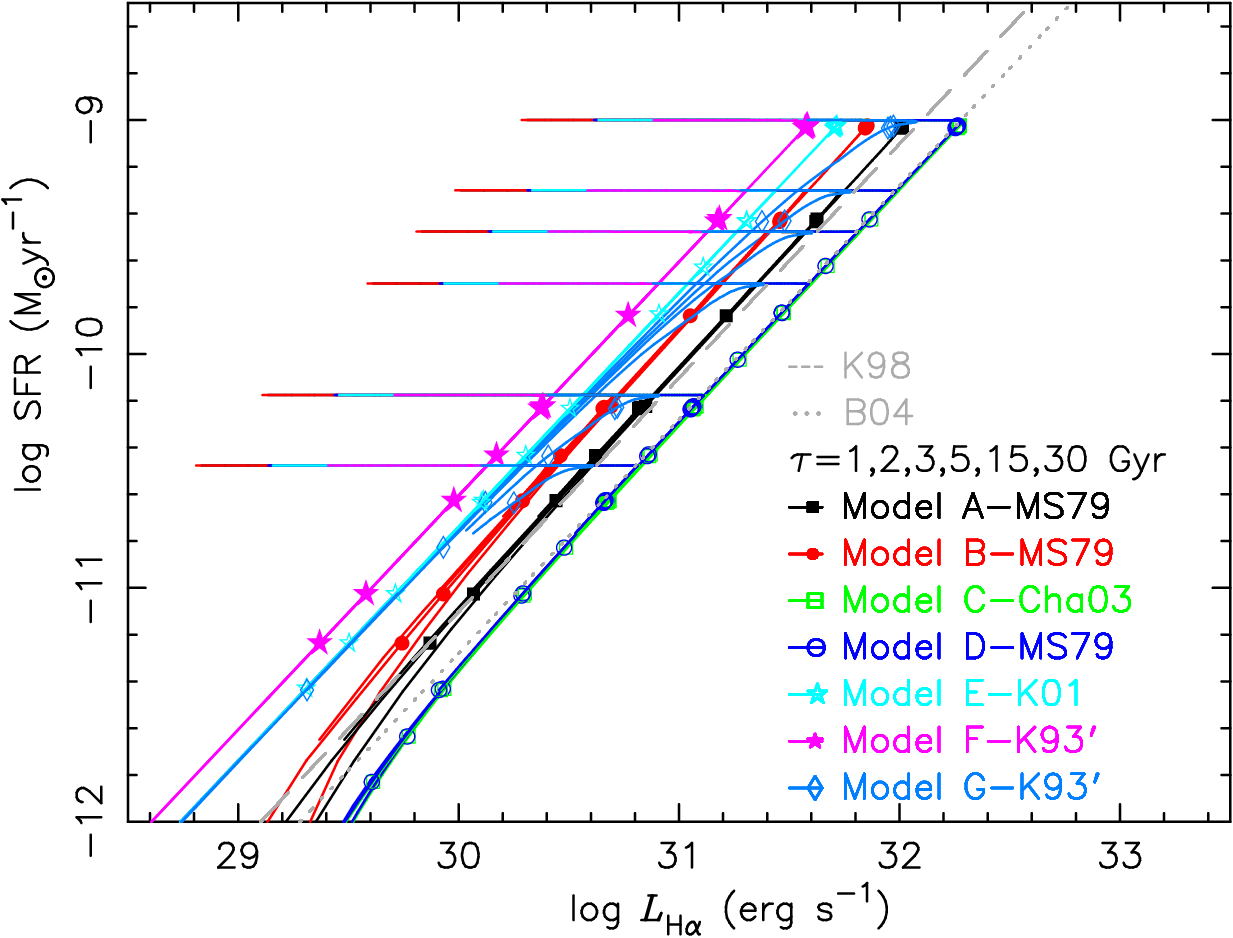}
\caption{Similar to Fig.~\ref{Fig:sfr-lha}. Left panel is for all Models using S55 IMF (including A-S55, B-S55, C-S55, E-S55', F-S55 and G-S55). Right panel is for all Models using NON-S55 IMFs (including A-MS79, B-MS79, C-Cha03, D-MS79, E-K01, F-K93' and G-K93').}
\label{Fig:sfr-lha-mod}
\end{figure*}

\begin{figure}
\centering
\includegraphics[height=8.0cm,width=6.5cm,angle=270]{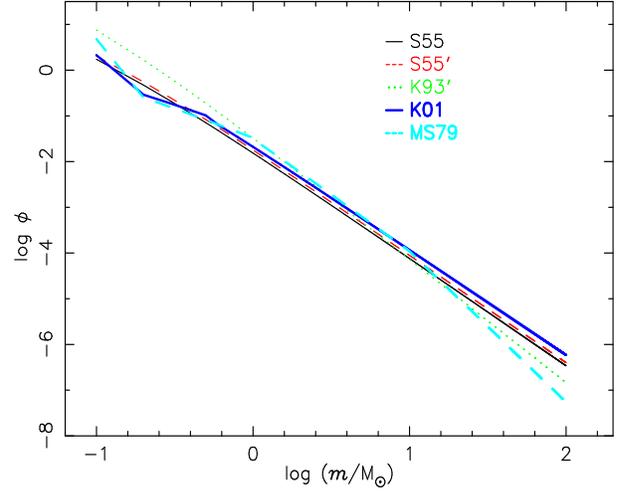}
\centering
\caption{Comparison of IMF (including S55, S55', K93', K01 and MS79). The detailed descriptions are presented in Section 2.}
\label{Fig:imf-all}
\end{figure}

\begin{figure*}
\includegraphics[bb=108 70 505 662,clip,angle=270,scale=.400]{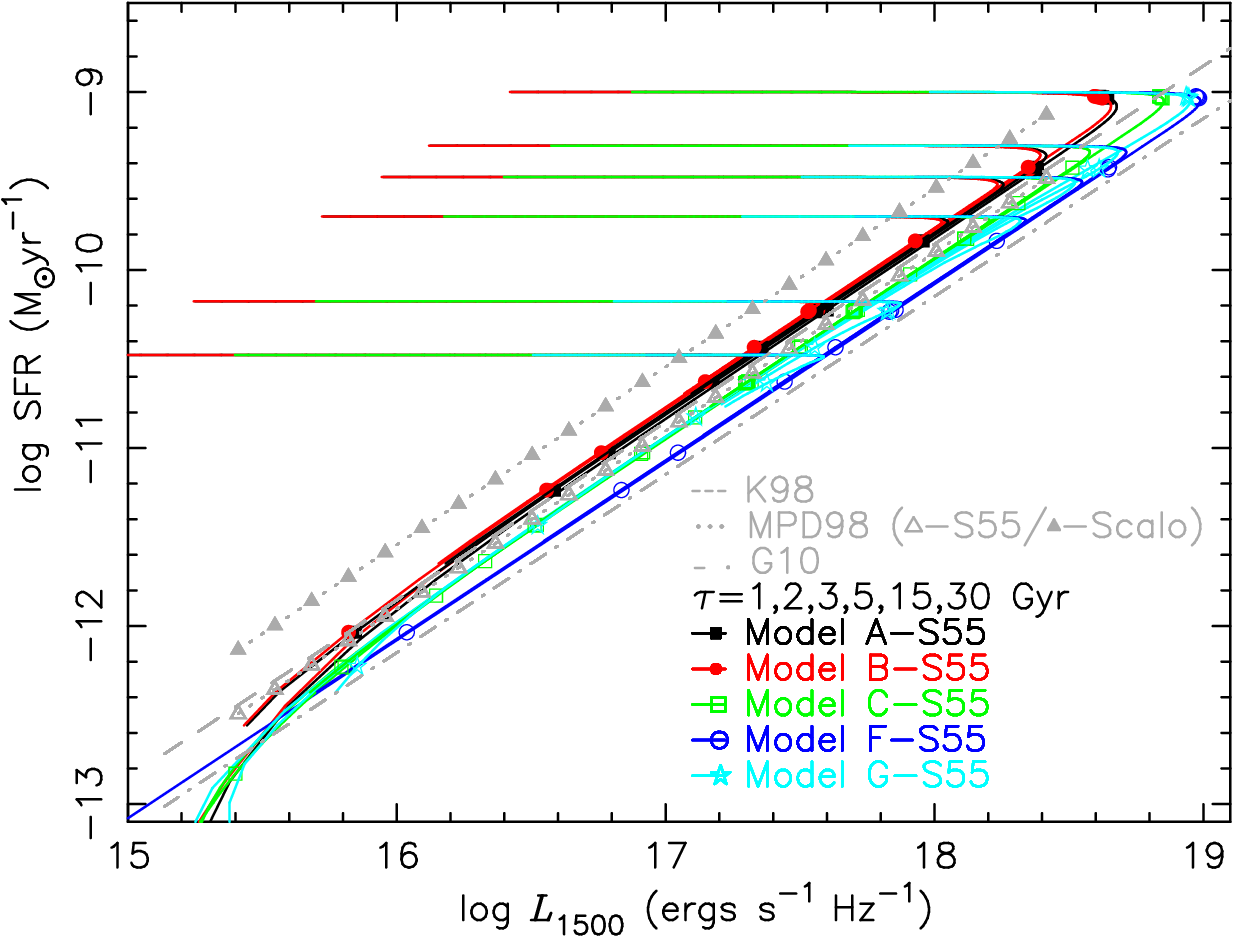}
\includegraphics[bb=108 70 505 662,clip,angle=270,scale=.400]{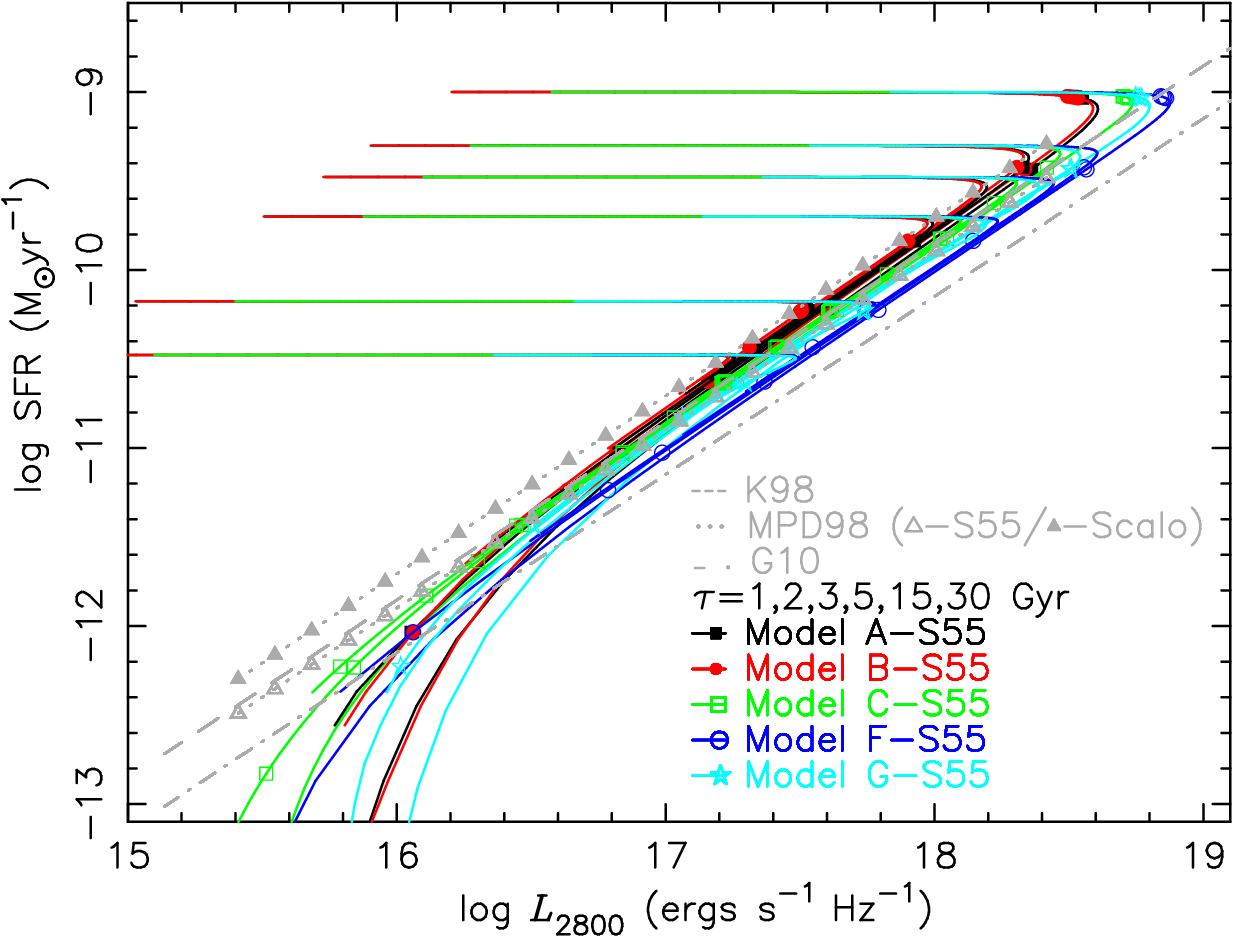}

\includegraphics[angle=270,scale=.400]{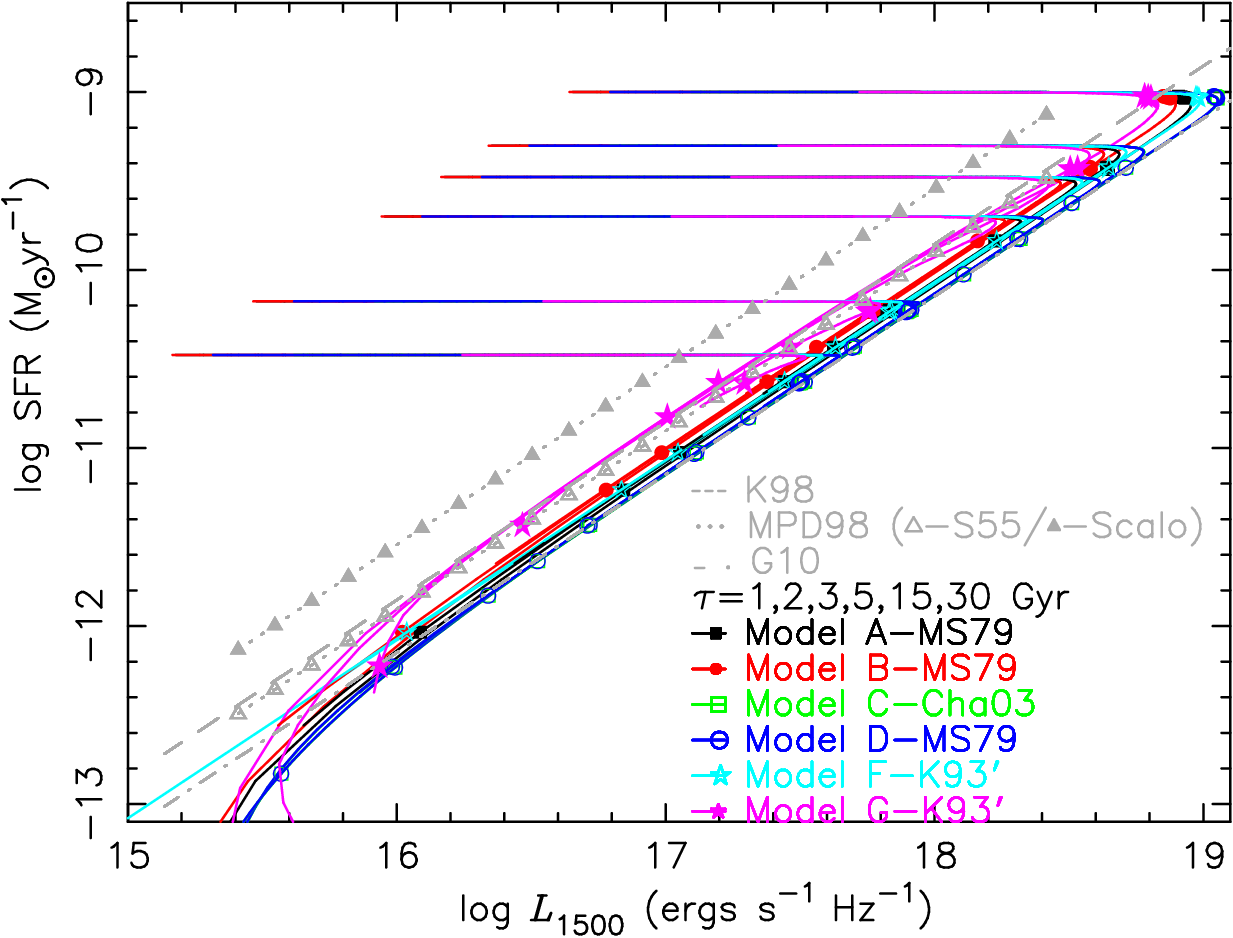}
\includegraphics[angle=270,scale=.400]{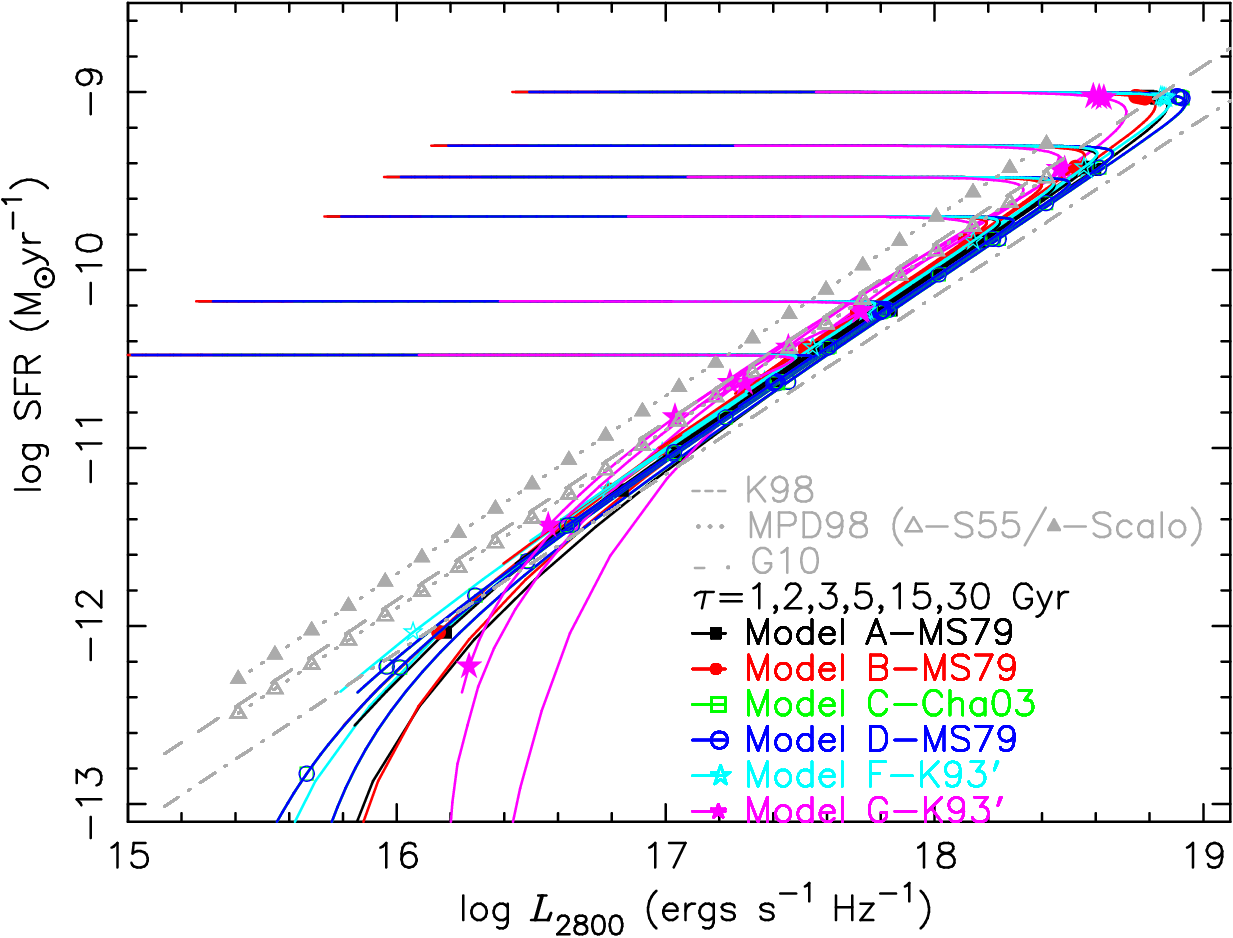}
\caption{Similar to Fig.~\ref{Fig:sfr-uv}. Left and right panels correspond to $L_{\rm 1500}$ and $L_{\rm 2800}$, respectively. Top panels represent all the models using S55 IMF (including A-S55, B-S55, C-S55, E-S55', F-S55 and G-S55). Bottom panels show all the models using NON-S55 IMFs (including A-MS79, B-MS79, C-Cha03, D-MS79, E-K01, F-K93' and G-K93').}
\label{Fig:sfr-uv-mod}
\end{figure*}

\begin{figure*}
\centering
\includegraphics[angle=270,scale=.400]{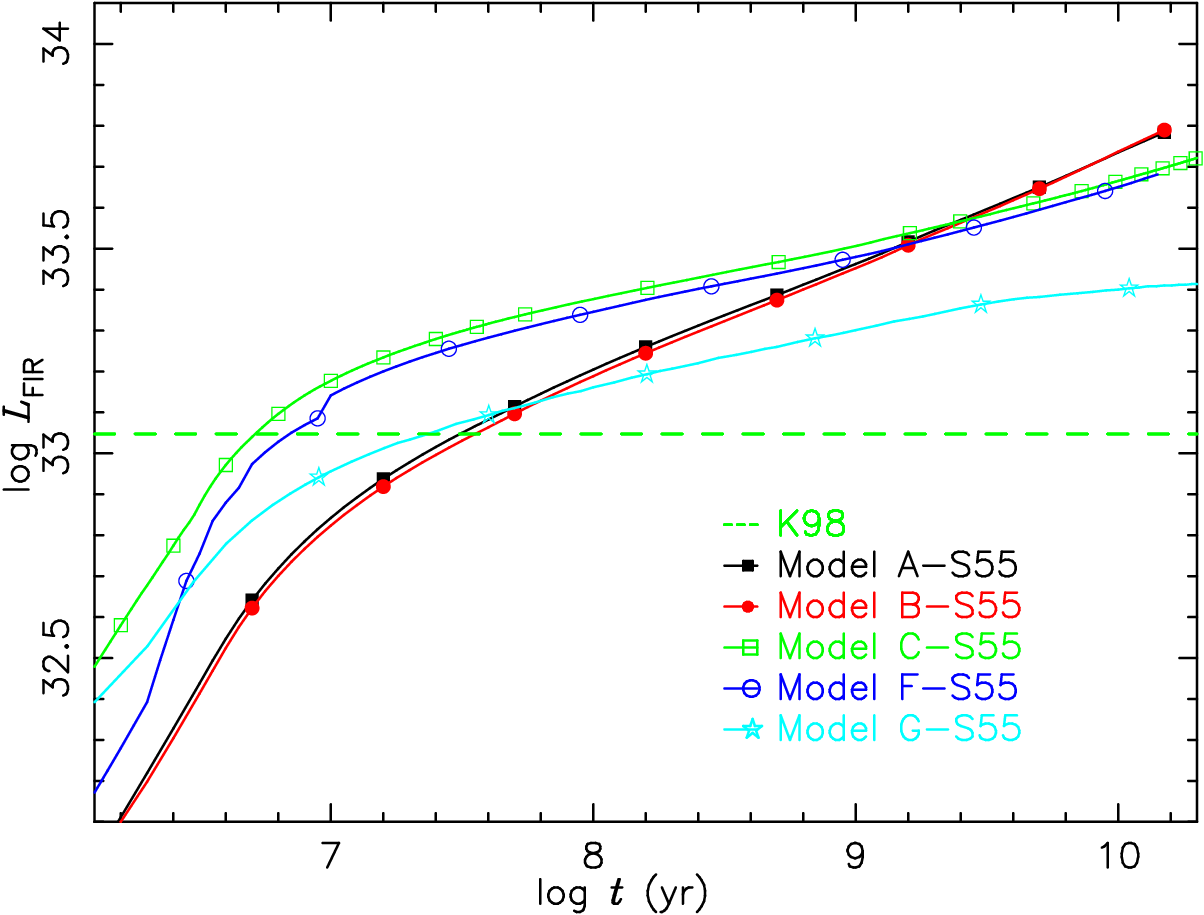}
\includegraphics[angle=270,scale=.400]{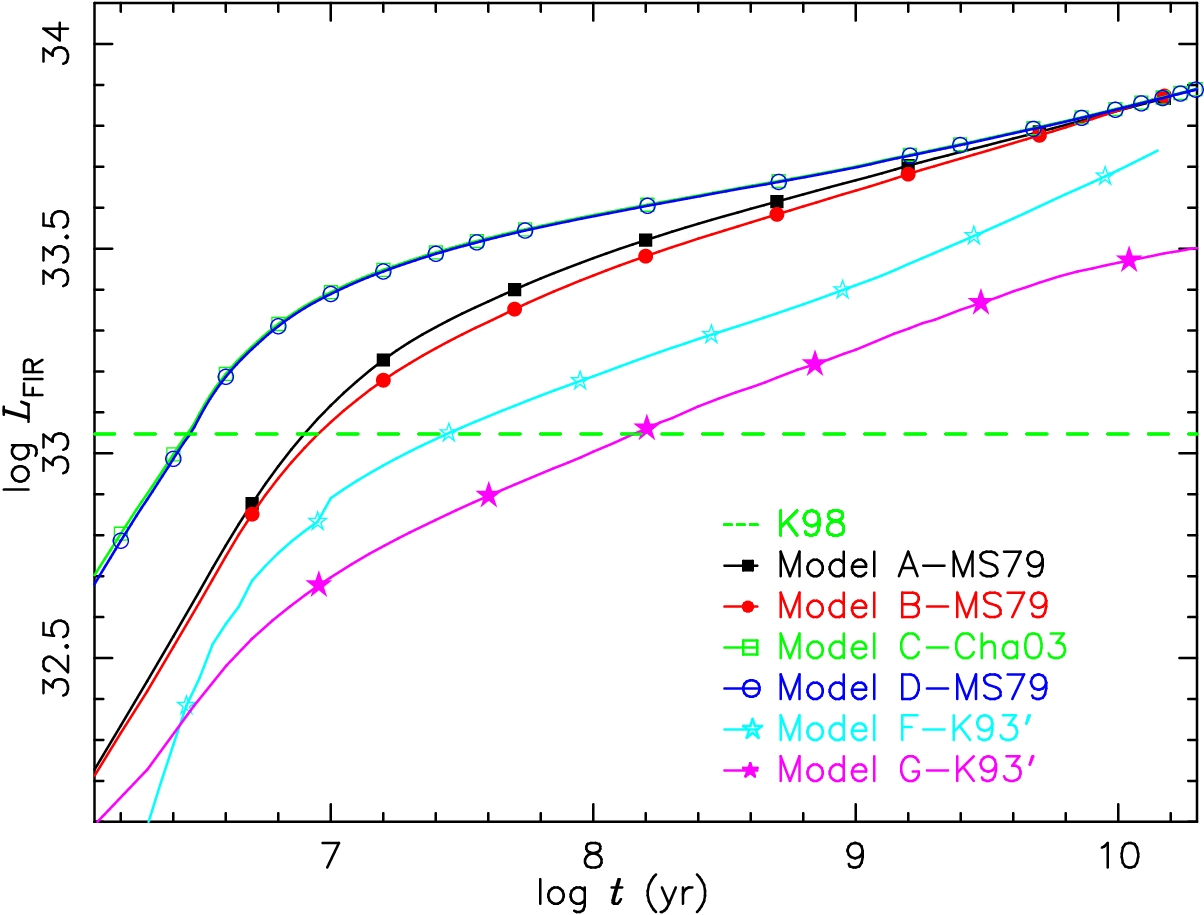}
\caption{Similar to Fig.~\ref{Fig:fir-t}. Left panel represents all the models using S55 IMF (including A-S55, B-S55, C-S55, E-S55', F-S55 and G-S55). Right panel shows all the models using NON-S55 IMFs (including A-MS79, B-MS79, C-Cha03, D-MS79, E-K01, F-K93' and G-K93').}
\label{Fig:Lfir-t-mod}
\end{figure*}

We use all Models, except for the set of Model Cr, to analyse the adoption of different EPS models on these $SFR$ calibrations. These models use the EPS models of {\hf Yunnan, GISSEL98, BC03, PopSTAR, STARBURST99} and {\hf P\'{E}GASE}, respectively.

To decrease the influence of IMF, we divide the comparisons into two groups. One group uses the S55 IMF and another group uses NON-S55 (including K01, K93', MS79 and Cha03) IMF. Moreover, we only compare our results with those studies by using the same (similar) IMF.

\subsubsection{EPS models on $SFR$.vs.$L_{\rm H\alpha}$}
In Fig.~\ref{Fig:sfr-lha-mod} we give the comparison in the $SFR$-$L_{\rm H\alpha}$ relation for all models by using S55 and NON-S55 IMFs.

At first, from the left panel of Fig.~\ref{Fig:sfr-lha-mod} we see that Models A-S55, B-S55 ({\hf Yunnan}) and F-S55 ({\hf STARBURST99}) give the similar $SFR$-$L_{\rm H\alpha}$ calibration and the corresponding curves lie above the line of K98.
The calibration curve of Model E-S55' ({\hf PopSTAR}) locates below the line of K98 and is the lowest one; the calibration curve of Model C-S55 ({\hf BC03}) is close to that of K98; and the calibration curve of Model G-S55 ({\hf P\'{E}GASE}) lie between Models C-S55 and E-S55' and does not display a linear calibration relation. The difference in the $SFR$.vs.$L_{\rm H\alpha}$ calibration caused by the adoption of different EPS models can reach $\sim $ 0.7\,dex when using the S55 IMF.

From the right panel of Fig.~\ref{Fig:sfr-lha-mod} we see that the calibration curves of Models C-Cha03 ({\hf BC03}) together with D-MS79 ({\hf GISSEL98}) overlap the line of B04, and the calibration curves of the other models locate above that of B04 (from bottom to up, A-MS79, B-MS79, E-K01. F-K93'). The difference in the $SFR$.vs.$L_{\rm H\alpha}$ calibration is $\sim $ 0.9\,dex when using NON-S55 IMF. This difference is partly caused by the difference in the IMF. In Fig.~\ref{Fig:imf-all} we give the comparison among the S55, S55', K01, K93' and MS79 IMFs.

\subsubsection{EPS models on $SFR$.vs.$L_{1500}$ and $SFR$.vs.$L_{2800}$}
In Fig.~\ref{Fig:sfr-uv-mod} we give the comparisons in the $SFR$-$L_{1500}$ and $SFR$-$L_{2800}$ relations for all models except for the set of Model E ({\hf PopSTAR}), for which the ISEDs are not provided. Top and bottom panels are for the results based on S55 and NON-S55 (including K01, K93', MS79 and Cha03) IMFs, respectively.

At first, from the top panels of Fig.~\ref{Fig:sfr-uv-mod}, we see that the $SFR$.vs.$L_{1500}$ and $SFR$.vs.$L_{2800}$ conversion factors of all models with the S55 IMF (A-S55, B-S55, C-S55, F-S55 and G-S55) are similar to those of K98 and MPD98-S55 (i.e., the results with the S55 IMF).
The differences in the $SFR$.vs.$L_{1500}$ and $SFR$.vs.$L_{2800}$ conversion factors, caused by the adoption of different EPS models, are less than $\sim$0.3\,dex when using the S55 IMF.

From the bottom panels of Fig.~\ref{Fig:sfr-uv-mod}, we see that all models with NON-S55 IMF (A-MS79, B-MS79, C-Cha03, D-MS79, F-K93' and G-K93') give small $SFR$-$L_{1500}$ and $SFR$-$L_{2800}$ conversion factors than the corresponding ones of MPD98-Scalo and similar values to those of G10 (for MPD98-Scalo and G10, the NON-S55 IMF is used).

The differences in the $SFR$.vs.$L_{1500}$ and $SFR$.vs.$L_{2800}$ conversion factors, caused by the adoption of different EPS models, can reach $\sim$0.2\,dex when using NON-S55 IMF. This is partly caused by the difference in the IMF.

\subsubsection{EPS models on $SFR$.vs.$L_{\rm [OII]}$}
The effect of EPS models on the $SFR$-$L_{\rm [OII]}$ relation is the same as that on the $SFR$-$L_{\rm H\alpha}$ relation, i.e., the difference in the $SFR$.vs.$L_{\rm [OII]}$ conversion factor can reach $\sim 0.7$\,dex when using the S55 IMF, and $\sim 0.9$\,dex when using the NON-S55 IMF.

\subsubsection{EPS models on $SFR$.vs.$L_{\rm FIR}$}
In Fig.~\ref{Fig:Lfir-t-mod} we give the $L_{\rm FIR}$ evolution of Irr-type galaxies (i.e., models with const SFR) when using the S55 and NON-S55 IMFs for all models except for the set of Model E ({\hf PopSTAR}).
From this figure we see that the difference in the EPS models can cause the difference of 0.4\,dex in the $SFR$.vs.$L_{\rm FIR}$ conversion factor when using the S55 IMF, and the difference of 0.8\,dex when using the NON-S55 IMF.

\section{Summary}
We use the {\hf Yunnan} EPS models with and without binary interactions to present the luminosities of the $\rm H\alpha$ recombination line, the [OII]$\lambda$3727 forbidden-line doublet, the UV (at $1500$ and $2800$\,$\rm \AA$) and the FIR continuum for Burst, E, S0, Sa-Sd and Irr galaxies, and present the calibrations of SFR in terms of these diagnostics.

By comparison, we find that binary interactions lower the SFR.vs.$L_{\rm H\alpha}$ and SFR.vs.$L_{\rm [OII]}$ conversion factors by $\sim $0.2\,dex, and do not significantly vary the SFR.vs.$L_{i,{\rm UV}}$ (at 1500 and 2800 $\rm \AA$) and SFR.vs.$L_{\rm FIR}$ calibrations.

We also consider the effects of IMF, the gas-recycle assumption and EPS models on these calibrations.
By comparison, we find that the SFR.vs.$L_{\rm H\alpha}$ and SFR.vs.$L_{\rm [OII]}$ conversion factors of Models A/B-S55 are larger by 0.4 and 0.2\,dex than the corresponding ones of Models A/B-MS79, and that the SFR.vs.$L_{i,{\rm UV}}$ and SFR.vs.$L_{\rm FIR}$ conversion factors are larger by 0.2\,dex.
By comparing the results between Models C and Cr, we find that the inclusion of gas-recycle assumption only lowers the SFR calibrations at faint SFR.
Also we use the other EPS models ({\hf BC03}, {\hf GISSEL98}, {\hf PopSTAR}, {\hf P\'{E}GASE} and {\hf STARBURST99}) to obtain these SFR calibrations. By comparison, we find that the differences in the SFR($L_{\rm H\alpha}$) and SFR($L_{\rm [OII]}$) calibrations reach $\sim$ 0.7 and 0.9\,dex, the difference in the SFR($L_{\rm FIR}$) calibration reaches 0.4 and 0.8\,dex, and the differences in the SFR($L_{i,{\rm UV}}$) calibration reach 0.3 and 0.2\,dex when using S55 and NON-S55 (partly caused by the difference in the IMF) IMFs, respectively.

At last, in this paper we give the conversion coefficients between $SFR$ and these diagnostics for all models.
In this paper we have only considered the effects of binary interactions for solar metallicity galaxies - more detailed studies will be given.

\section*{acknowledgements}
This work was funded by the Chinese Natural Science Foundation (Grant Nos 10773026, 11073049, 11033008, 10821026 \& 2007CB15406), by the Yunnan Natural Science Foundation (Grant No 2007A113M) and by the Chinese Academy of Sciences (KJCX2-YW-T24). We are also grateful to the referee for suggestions that have improved the quality of this manuscript.



\bsp
\label{lastpage}
\end{document}